\documentclass[a4paper,journal]{IEEEtran}
\usepackage{amsmath,amsfonts}
\usepackage{algorithmic}
\usepackage{algorithm}
\usepackage{array}
\usepackage[caption=false,font=normalsize,labelfont=sf,textfont=sf]{subfig}
\usepackage{textcomp}
\usepackage{stfloats}
\usepackage{url}
\usepackage{verbatim}
\usepackage{graphicx}
\usepackage{cite}
\usepackage{multirow}
\usepackage{multicol}
\usepackage{booktabs}
\begin{document}

\title{Amortized Neural Optimization for Pre-Layout Signal Integrity Design Space Exploration using Differentiable Surrogates}

\author{Julian Withöft, Werner John, Emre Ecik, Ralf Brüning, Jürgen Götze
\thanks{This work has been submitted to the IEEE for possible publication. Copyright may be transferred without notice, after which this version may no longer be accessible.}%
\thanks{This article is an extended version of a paper presented at the 2026 IEEE International Symposium on Electromagnetic Compatibility and Signal/Power Integrity (EMC+SIPI) \cite{Withoeft2026}.}%
\thanks{This work is funded as part of the research project KI4BoardNet in the funding programme MANNHEIM (BMBF) (Grant numbers 16ME0779/TUDO - 16ME0777/IDMT). The responsibility for this publication is held by the authors only.}%
\thanks{Julian Withöft, Emre Ecik and Jürgen Götze are with the Information Processing Lab, Faculty for Electrical Engineering and Information Technology, TU Dortmund, Germany (e-mail: \{julian.withoeft, emre.ecik, juergen.goetze\}@tu-dortmund.de)}%
\thanks{Werner John is with Pyramide2525, Paderborn, Germany and the Information Processing Lab, Faculty for Electrical Engineering and Information Technology, TU Dortmund, Germany (e-mail: werner.john@pyramide2525.de)}%
\thanks{Ralf Brüning is with EMC Technology Center Paderborn, Zuken GmbH, Paderborn, Germany (e-mail: ralf.bruening@de.zuken.com)}%
}

\maketitle

\begin{abstract}
Pre-layout design space exploration (DSE) for high-speed signal integrity (SI) analysis is often limited by the computational cost of simulations and iterative optimization algorithms within modern electronic design automation (EDA) workflows. While machine learning surrogate models accelerate the simulation step, optimizing designs still requires utilizing iterative black-box search methods. This iterative nature scales poorly and makes multi-corner sweeps computationally expensive. As a solution, this paper proposes amortized neural optimization (ANO) for pre-layout SI design. ANO avoids iterative black-box inference and utilizes a fully differentiable neural network surrogate model from which it extracts gradient information to train a global optimization policy. This does not solve the optimization problem repeatedly at inference, but learns the process offline, which amortizes the computational cost. Once the ANO policy is trained, it maps different channel contexts to near-optimal design parameters in a single deterministic forward pass. The efficiency and accuracy of the ANO framework are demonstrated based on three SI design scenarios, including DDR5 decision feedback equalization (DFE), 9-dimensional SerDes Tx/Rx co-equalization, and DDR3 DQS differential pair eye diagram optimization under intra-pair skew constraints. Trading roughly 10\% in optimality compared to instance-specific black-box algorithms results in speedups of three to four orders of magnitude. For a large-scale 320,000-instance multi-corner SerDes sweep optimization, ANO reduces what would have taken days of computation using iterative search algorithms to a single batched forward pass that completes in milliseconds. This transforms computationally expensive SI optimization into real-time and interactive pre-layout DSE.
\end{abstract}


\section{Introduction}

\IEEEPARstart{P}{re-layout} design space exploration (DSE) for signal integrity (SI) analysis on high-speed interconnects plays an essential role within modern electronic design automation (EDA) workflows to achieve reliable printed circuit board (PCB) layouts. During this phase, key decisions regarding critical high-speed nets are made, which makes it essential to obtain fast and reliable feedback on which configurations will satisfy target SI specifications.

The pre-layout design phase provides the greatest flexibility in the overall design process, as fundamental architectural decisions can still be adjusted with comparatively low cost and effort. Therefore, it represents the stage with the highest degree of design freedom and the greatest impact on overall system performance. However, evaluating the large number of possible designs through conventional simulation-based approaches is time-consuming and computationally expensive. Moreover, it requires significant domain expertise, making the process challenging, especially for less experienced designers.

In this context, machine learning (ML) methods have great potential in the form of artificial intelligence (AI)-assisted decision support and design assistance EDA tools. By training ML models on large amounts of simulation data, they provide fast SI performance predictions during the pre-layout design exploration phase. This allows designers to analyze the design space much more efficiently and to identify promising configurations early, while also reducing the number of costly design iterations to support SI-compliant PCB development.

For ML-based SI design, significant effort has been focused on the prediction of eye diagram metrics such as eye-height (EH) and eye-width (EW) from PCB design parameters. Different forward modeling approaches have been applied, including neural networks (NNs) \cite{Lu2018, Zhang2022, Withoeft2024, Withoeft2025}, support vector regression (SVR) \cite{Lu2018, Ma2022}, and Gaussian process regression (GPR) \cite{Nguyen2023}. Other approaches utilized NNs and kernel ridge regression to predict entire inner eye contours for direct mask-compliance checking \cite{Goay2019, Telescu2023}. However, while these forward surrogates accelerate the SI evaluation step, they do not solve the design optimization problem. Identification of optimal PCB geometries or equalizer settings still requires combining the surrogate with an iterative search algorithm.

Historically, SI design optimization has relied on iterative black-box search methods applied either directly based on simulators or using ML surrogate models. For direct simulation-based tuning, Bayesian optimization (BO) is frequently utilized, for example, for continuous-time linear equalization (CTLE) \cite{Bohl2023}, transmitter-side (Tx) feedforward equalization (FFE) \cite{Dolatsara2022}, and combined receiver-side (Rx) FFE and decision feedback equalization (DFE) tuning \cite{Kiguradze2020, Dikhaminjia2021}. While BO is very sample-efficient for expensive black-box simulations, there is a contradiction when BO is paired with a fast ML surrogate model, as the probabilistic acquisition function and the underlying Gaussian process introduce mathematical overhead that completely negates the speed of the underlying ML surrogate.

In contrast to this surrogate incompatibility of BO, evolutionary algorithms such as genetic algorithms (GAs) \cite{Zhang2022, Withoeft2024, Withoeft2025} and differential evolution (DE) \cite{Sun2024} have been utilized directly in combination with ML surrogates to optimize EH/EW metrics or Tx-side FFE settings. To further reduce search times, hybrid approaches, such as combining SVR surrogates with active-subspace reduction, have been proposed to constrain the search space for FFE, CTLE, and DFE tuning \cite{Ma2022}. Even though the population-based methods like GAs are faster compared to BO because they can evaluate batches of candidates simultaneously on the surrogate, they are still iterative. As a consequence, for large-scale DSE that requires the evaluation of thousands of varying channel configurations, the total runtime of the iterative loop is still large. More critically, all of these methods treat the surrogate model just as a fast black-box oracle and fail completely to exploit the continuous differentiability and gradients of NN architectures.

To avoid iterative search completely, inverse modeling approaches attempt to map desired SI targets back to the required PCB and equalizer parameters. Direct NN- and SVR-based surrogate models targeted specific transient response metrics such as overshoot and rise time \cite{Withoeft2023a, Withoeft2023b}, or eye diagram metrics \cite{Trinchero2019, Roy2019}. However, standard direct modeling experiences problems with the ambiguity of SI design, where multiple different combinations of design parameters can result in identical SI metrics. Specialized invertible neural networks (INNs) are particularly effective to address this ambiguity and one-to-many mapping challenge. The basic framework was implemented in \cite{Dolatsara2020}, which demonstrated INNs based on two CTLE parameters and was later extended in \cite{Ambasana2021} to include a total of five parameters with additional PCB variables like trace length and impedance. Benchmarking in \cite{Bhatti2021} has also demonstrated that INNs are superior in comparison to direct feedforward inverse NNs and conditional generative adversarial networks (cGANs). However, all inverse models share a limitation as they rely on pre-specified targets. This forces the designer into a guessing game of either risking to request physically infeasible metrics or under-utilizing the channel's maximum capacity. Additionally, their probabilistic properties lead to a distribution of potential designs rather than a single deterministic optimal solution, which requires subsequent selection and simulation verification.

To achieve near-instantaneous inference without requiring pre-defined targets, recent research has moved towards reinforcement learning (RL) and imitation learning. RL has been applied to learn global SI optimization policies for tasks such as double data rate (DDR) eye aperture maximization \cite{Lho2022}, hybrid equalizer optimization for high-bandwidth memory (HBM) modules \cite{Choi2023}, and optimization of transient response parameters in DDR memory systems \cite{Withoeft2023c}. In contrast, imitation learning has been utilized to train neural networks that mimic the behavior of expert black-box optimizers such as GA or BO for reusable FFE/CTLE optimization \cite{Kim2023a, Choi2024} and dimensioning of vias \cite{Kim2023b}.

However, while the inference is near-instantaneous, training these policy networks requires large datasets from computationally intensive simulations. Imitation learning is especially constrained by this requirement, as the expert datasets must be generated by the same iterative search algorithms, such as GA or BO, that were previously established as computationally inefficient, which causes the method to inherit their computational overhead. Concurrently, RL suffers from sample inefficiency and the required empirical and fragile reward shaping process. Additionally, there is an architectural mismatch, as RL is designed for complex sequential state transitions and Markov processes rather than the bandit-like single-step mapping of a channel context to its optimal design parameters. Finally, because both methods train neural networks to obtain global optimization policies based on trial-and-error black-box interactions without access to gradients to guide the optimization, achieving convergence is a resource-intensive and sample-inefficient process.

Recent methodologies in S-parameter modeling have demonstrated that utilizing the continuous differentiability of neural surrogates allows for direct gradient-based optimization with respect to the design parameters \cite{Akinwande2023}. Gradient descent is more sample-efficient than black-box search or RL, but it still is a local iterative procedure at core. For optimization, it requires repeated forward and backward passes through the neural network for every new instance. Moreover, this method is susceptible to getting stuck in local minima, which results in sub-optimal results from a global point of view.

To overcome these limitations of iterative search algorithms, inverse modeling, and black-box policy learning, this paper proposes an amortized neural optimization (ANO) framework that utilizes differentiable surrogates for pre-layout SI-DSE. Building upon the proof-of-concept introduced in \cite{Withoeft2026}, this extended work benchmarks the ANO framework against a comprehensive suite of state-of-the-art optimization methods, which demonstrates that the architecture generalizes to different complex SI design tasks, including co-equalization and constrained PCB parameter optimization.

This method uses a dual-network setup with a differentiable NN surrogate model and a global ANO policy network. The ANO framework does not treat SI analysis as a black-box problem, but utilizes gradient descent directly through the frozen surrogate \cite{Sun2021} to update the weights of the ANO policy network. With this gradient information, the ANO policy can be trained for minimization of a specific loss function, which is based on a custom objective function for the given optimization scenario. This avoids guessing fixed a priori performance targets like in inverse modeling and also provides a much more guided optimization path compared to the fragile RL reward shaping process.

While iterative black-box algorithms such as GA or BO applied directly based on simulators may be efficient for tuning a single isolated channel instance, they become computationally expensive for large-scale DSE. Even when accelerated with fast ML surrogates, their sequential nature re-introduces a computational bottleneck that must be repeated from scratch for every new design instance, while the ANO framework overcomes this limitation. In alignment with the general principles of amortized optimization \cite{Amos2023}, the computational complexity of the optimization process is amortized offline during the training phase and, during inference, the learned policy maps channel contexts to optimized deterministic design configurations in a single forward pass. This transforms design space sweeps that would have taken multiple days using evolutionary search algorithms into a near-instantaneous process that finishes within milliseconds. To demonstrate this efficiency and also the applicability, the ANO methodology is evaluated across diverse high-speed SI scenarios. To realize these SI optimization tasks, advanced techniques are introduced, such as multi-task learning for shared multi-receiver equalization and curriculum penalties for constrained layout optimization.

\section{Signal Integrity Application Examples}

\begin{figure}[b!]
	\centering
	\includegraphics[width=8.5cm]{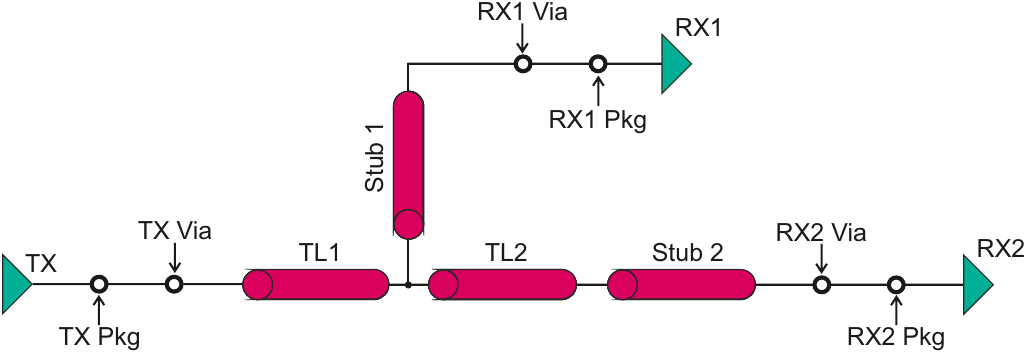}
	\caption{DDR5 daisy chain PCB structure with two receivers \cite{Withoeft2026}.}
	\label{fig_ddr5_interconnect}
\end{figure}

To demonstrate the ANO approach and its feasibility for fast and efficient pre-layout SI design space exploration, three different SI application scenarios were considered.

\subsection{DDR5 Daisy Chain DFE Equalization Optimization}

\begin{table}[b!]
	\centering
	\caption{Parameter ranges for the DDR5 daisy chain \cite{Withoeft2026}}
	\label{tab_params_ddr5}
	\begin{tabular}{lcc}
		\toprule
		\textbf{Parameter [Unit]} & \textbf{Min.} & \textbf{Max.} \\
		\midrule
		Clock Frequency $f_\mathrm{clock}$ [GHz] & 2.4 & 3.6 \\
		Supply Voltage $V_\mathrm{DDQ}$ [V] & 1.0 & 1.4 \\
		Jitter [\% of $T_\mathrm{clock}$] & 0 & 3 \\
		Rise Time [\% of $T_\mathrm{clock}$] & 10 & 25 \\
		Fall Time [\% of $T_\mathrm{clock}$] & 10 & 25 \\
		Tx Resistor $R_\mathrm{Tx}$ [$\Omega$] & 30 & 50 \\
		Tx Capacitance $C_\mathrm{Tx}$ [pF] & 0.1 & 2.0 \\
		Tx Package Resistor $R_\mathrm{pkg,Tx}$ [$\Omega$] & 0.01 & 0.50 \\
		Tx Package Inductor $L_\mathrm{pkg,Tx}$ [nH] & 0.1 & 1.5 \\
		Tx Package Capacitance $C_\mathrm{pkg,Tx}$ [pF] & 0.1 & 0.8 \\
		Tx Via Resistor $R_\mathrm{via,Tx}$ [$\Omega$] & 0.001 & 0.10 \\
		Tx Via Inductor $L_\mathrm{via,Tx}$ [nH] & 0.1 & 0.8 \\
		Tx Via Capacitance $C_\mathrm{via,Tx}$ [pF] & 0.1 & 0.4 \\
		\midrule
		Transmission Line 1 Length $L_\mathrm{TL1}$ [mm] & 10 & 120 \\
		Transmission Line 1 Impedance $Z_{0,\mathrm{TL1}}$ [$\Omega$] & 30 & 70 \\
		Transmission Line 1 Dielectric $\epsilon_{r,\mathrm{TL1}}$ & 3 & 5 \\
		Transmission Line 1 Loss Tangent $\tan\delta_\mathrm{TL1}$ & 0.003 & 0.045 \\
		Stub 1 Length $L_\mathrm{Stub1}$ [mm] & 0.1 & 10 \\
		Stub 1 Impedance $Z_{0,\mathrm{Stub1}}$ [$\Omega$] & 30 & 70 \\
		Stub 1 Dielectric $\epsilon_{r,\mathrm{Stub1}}$ & 3 & 5 \\
		Stub 1 Loss Tangent $\tan\delta_\mathrm{Stub1}$ & 0.003 & 0.045 \\
		\midrule
		Rx1 Via Resistor $R_\mathrm{via,Rx1}$ [$\Omega$] & 0.001 & 0.10 \\
		Rx1 Via Inductor $L_\mathrm{via,Rx1}$ [nH] & 0.1 & 0.8 \\
		Rx1 Via Capacitance $C_\mathrm{via,Rx1}$ [pF] & 0.1 & 0.4 \\
		Rx1 Package Resistor $R_\mathrm{pkg,Rx1}$ [$\Omega$] & 0.01 & 0.50 \\
		Rx1 Package Inductor $L_\mathrm{pkg,Rx1}$ [nH] & 0.1 & 1.5 \\
		Rx1 Package Capacitance $C_\mathrm{pkg,Rx1}$ [pF] & 0.1 & 0.8 \\
		Rx1 Input Capacitance $C_\mathrm{Rx1}$ [pF] & 0.1 & 2.0 \\
		Rx1 ODT Termination $R_\mathrm{odt,Rx1}$ [$\Omega$] & 30 & 120 \\
		Rx1 DFE Tap 1 & -0.40 & 0.10 \\
		Rx1 DFE Tap 2 & -0.15 & 0.15 \\
		Rx1 DFE Tap 3 & -0.10 & 0.10 \\
		Rx1 DFE Tap 4 & -0.08 & 0.08 \\
		\midrule
		Transmission Line 2 Length $L_\mathrm{TL2}$ [mm] & 10 & 80 \\
		Transmission Line 2 Impedance $Z_{0,\mathrm{TL2}}$ [$\Omega$] & 30 & 70 \\
		Transmission Line 2 Dielectric $\epsilon_{r,\mathrm{TL2}}$ & 3 & 5 \\
		Transmission Line 2 Loss Tangent $\tan\delta_\mathrm{TL2}$ & 0.003 & 0.045 \\
		Stub 2 Length $L_\mathrm{Stub2}$ [mm] & 0.1 & 10 \\
		Stub 2 Impedance $Z_{0,\mathrm{Stub2}}$ [$\Omega$] & 30 & 70 \\
		Stub 2 Dielectric $\epsilon_{r,\mathrm{Stub2}}$ & 3 & 5 \\
		Stub 2 Loss Tangent $\tan\delta_\mathrm{Stub2}$ & 0.003 & 0.045 \\
		\midrule
		Rx2 Via Resistor $R_\mathrm{via,Rx2}$ [$\Omega$] & 0.001 & 0.10 \\
		Rx2 Via Inductor $L_\mathrm{via,Rx2}$ [nH] & 0.1 & 0.8 \\
		Rx2 Via Capacitance $C_\mathrm{via,Rx2}$ [pF] & 0.1 & 0.4 \\
		Rx2 Package Resistor $R_\mathrm{pkg,Rx2}$ [$\Omega$] & 0.01 & 0.50 \\
		Rx2 Package Inductor $L_\mathrm{pkg,Rx2}$ [nH] & 0.1 & 1.5 \\
		Rx2 Package Capacitance $C_\mathrm{pkg,Rx2}$ [pF] & 0.1 & 0.8 \\
		Rx2 Input Capacitance $C_\mathrm{Rx2}$ [pF] & 0.1 & 2.0 \\
		Rx2 ODT Termination $R_\mathrm{odt,Rx2}$ [$\Omega$] & 30 & 120 \\
		Rx2 DFE Tap 1 & -0.40 & 0.10 \\
		Rx2 DFE Tap 2 & -0.15 & 0.15 \\
		Rx2 DFE Tap 3 & -0.10 & 0.10 \\
		Rx2 DFE Tap 4 & -0.08 & 0.08 \\
		\bottomrule
	\end{tabular}
\end{table}

Firstly, a simplified DDR5 daisy chain interconnect with two receivers (Rx1/Rx2) \cite{jedec}, shown in Fig.~\ref{fig_ddr5_interconnect}, was considered \cite{Withoeft2026}. The simulation setup utilized a generalized linear transmitter (Tx) model parameterized in accordance with \cite{jedec}, including clock frequency, supply voltage, rise/fall times, internal resistance and capacitance, as well as jitter. In addition, package and via parasitics were incorporated, and the interconnect traces and stubs were represented as frequency-dependent lossy transmission lines using multi-segment RLGC ladder networks.

At each receiver, on-die termination (ODT), input capacitance, and a 4-tap decision feedback equalizer (DFE) were included to account for receiver-side equalization effects. ODT was treated as a static tunable parameter at both receivers to provide a generalized multi-drop termination abstraction for optimization. However, the proposed ANO framework remains directly applicable to typical far-end ODT and multi-DRAM fly-by configurations. The simulations were performed using LTspice \cite{LTspice} with a PRBS-10 excitation. The DFE was subsequently applied in a dedicated post-processing step to the simulated waveform in order to compute the final post-DFE worst-case inner eye contour, which was evaluated at $\mathrm{SL}$ = 50 sampling points within the unit interval (UI). This resulted in a 100-dimensional output vector, comprising 50 samples for the upper contour and 50 samples for the lower contour.

A total of 53 design parameters were varied to generate the dataset for training the ML surrogate model. Among these, there were 45 PCB-related context parameters $C$, including the transmitter, vias, packages, interconnects, and receivers. The remaining 8 tunable parameters $T$ corresponded to the DFE tap equalization parameters, with four taps assigned to each receiver. Both sets of parameters, $C$ and $T$, were varied according to the parameter ranges in Table~\ref{tab_params_ddr5} using Latin hypercube sampling (LHS) to produce a dataset consisting of 40,000 training, 10,000 validation, and 10,000 testing samples.

\subsection{SerDes Co-Equalization Optimization}

\begin{table}[b!]
	\centering
	\caption{Parameter ranges for the SerDes interconnect}
	\label{tab_params_serdes}
	\begin{tabular}{lcc}
		\toprule
		\textbf{Parameter [Unit]} & \textbf{Min.} & \textbf{Max.} \\
		\midrule
		Data Rate [Gbps] & 10 & 32 \\
		Differential Swing $V_\mathrm{swing}$ [V] & 0.4 & 1.2 \\
		Tx Jitter [\% of UI] & 0.1 & 3.0 \\
		Tx Rise/Fall Times [\% of UI] & 10 & 40 \\
		\midrule
		FFE (Tx) Pre-Cursor Tap & -0.15 & 0.0 \\
		FFE (Tx) Post-Cursor Tap & -0.35 & 0.0 \\
		CTLE (Rx) Boost Magnitude [dB] & 0 & 15 \\
		CTLE (Rx) Pole 1 Ratio [$f_{p1}/f_\mathrm{Nyq}$] & 0.30 & 1.50 \\
		CTLE (Rx) Pole 2 Gap Ratio [$\Delta f_{p2}/f_\mathrm{Nyq}$] & 0.10 & 5.00 \\
		DFE (Rx) Tap 1 & -0.30 & 0.30 \\
		DFE (Rx) Tap 2 & -0.15 & 0.15 \\
		DFE (Rx) Tap 3 & -0.10 & 0.10 \\
		DFE (Rx) Tap 4 & -0.08 & 0.08 \\
		\midrule
		Tx Resistor $R_\mathrm{Tx}$ [$\Omega$] & 40 & 60 \\
		Tx Capacitance $C_\mathrm{Tx}$ [pF] & 0.1 & 0.3 \\
		Rx Resistor $R_\mathrm{Rx}$ [$\Omega$] & 40 & 60 \\
		Rx Capacitance $C_\mathrm{Rx}$ [pF] & 0.1 & 0.3 \\
		Tx Package Inductor $L_\mathrm{pkg,Tx}$ [nH] & 0.1 & 0.8 \\
		Tx Package Capacitance $C_\mathrm{pkg,Tx}$ [pF] & 0.1 & 0.4 \\
		Rx Package Inductor $L_\mathrm{pkg,Rx}$ [nH] & 0.1 & 0.8 \\
		Rx Package Capacitance $C_\mathrm{pkg,Rx}$ [pF] & 0.1 & 0.4 \\
		Rx RMS Noise [mV] & 0.5 & 3.0 \\
		\midrule
		Tx Via Inductor $L_\mathrm{via,Tx}$ [nH] & 0.1 & 0.6 \\
		Tx Via Pad Capacitance $C_\mathrm{pad,Tx}$ [pF] & 0.10 & 0.25 \\
		Tx Via Anti-pad Capacitance $C_\mathrm{anti,Tx}$ [pF] & 0.05 & 0.15 \\
		Rx Via Inductor $L_\mathrm{via,Rx}$ [nH] & 0.1 & 0.6 \\
		Rx Via Pad Capacitance $C_\mathrm{pad,Rx}$ [pF] & 0.10 & 0.25 \\
		Rx Via Anti-pad Capacitance $C_\mathrm{anti,Rx}$ [pF] & 0.05 & 0.15 \\
		\midrule
		Channel Length $L$ [mm] & 10 & 300 \\
		Trace Width $W$ [$\mu$m] & 70 & 300 \\
		Trace Spacing $S$ [$\mu$m] & 70 & 300 \\
		Substrate Height $H$ [$\mu$m] & 70 & 250 \\
		Dielectric Constant $\epsilon_r$ ($D_k$) & 3.0 & 5.0 \\
		Loss Tangent $\tan\delta$ ($D_f$) & 0.002 & 0.030 \\
		\bottomrule
	\end{tabular}
\end{table}

\begin{figure}[b!]
	\centering
	\includegraphics[width=8.5cm]{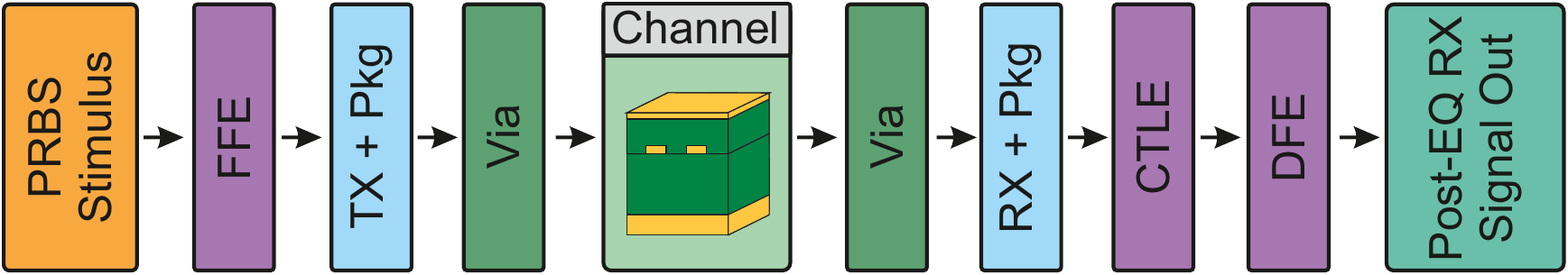}
	\caption{SerDes interconnect architecture.}
	\label{fig_serdes_interconnect}
\end{figure}

Secondly, the SerDes interconnect architecture shown in Fig.~\ref{fig_serdes_interconnect} with a Tx-side FFE, an Rx-side CTLE, and an Rx-side DFE was analyzed. To generate training data, transient simulations with a PRBS-16 stimulus truncated to 50,000 bits were performed in Matlab \cite{Matlab}. A 3-tap FFE was applied at the Tx to pre-compensate for intersymbol interference (ISI). This signal was then converted into a differential waveform by the transmitter based on configurable rise and fall times, differential swing, and jitter. The Tx signal then traveled through the transmitter package, vias, and a coupled stripline channel before reaching the receiver package and vias. At the receiver, the CTLE amplified frequency components attenuated by the channel to compensate for high-frequency loss. This was followed by a 4-tap DFE to cancel remaining post-cursor ISI.

A total of 34 design parameters, including stimulus characteristics, Tx/Rx equalization parameters, analog front-ends, via and package parasitics, and coupled stripline channel geometry, were varied within the ranges summarized in Table~\ref{tab_params_serdes}, while the trace thickness was kept constant at 17~µm (0.5 oz copper). LHS was used to generate datasets comprising 50,000 samples for training, 10,000 for validation, and 10,000 for testing. Each sample led to an eye diagram and the corresponding 1\% probability inner eye contour, which was evaluated at $\mathrm{SL}$ = 64 samples per UI. This resulted in a 128-dimensional output vector with 64 samples for the upper contour and 64 samples for the lower contour. In this scenario, there were 25 context parameters $C$ describing the physical SerDes channel design as well as Tx/Rx parameters and 9 tunable equalization parameters $T$ determining the active effect of the FFE, CTLE, and DFE.

\subsection{DDR3 DQS Eye Optimization under Skew Constraints}

\begin{table}[b!]
	\centering
	\caption{Parameter ranges for the DSSTL DDR3 DQS interconnect \cite{Withoeft2024}}
	\label{tab_params_ddr3}
	\begin{tabular}{lcc}
		\toprule
		\textbf{Parameter} & \textbf{Min.} & \textbf{Max.} \\
		\midrule
		$R_T = R_1 = R_2$ [$\Omega$] & 0 & 180 \\
		Length $DL_2$ [mm] & 1 & 200 \\
		Length $DL_1$ [mm] & 1 & 200 \\
		Length $DL_3$ [mm] & 1 & 200 \\
		Length $TL_3$ [mm] & 0.1 & 50 \\
		Trace width $W$ [µm] & 70 & 300 \\
		Diff. pair spacing $S$ [µm] & 70 & 3500 \\
		Dielectric constant $\epsilon_r$ & 4 & 4.8 \\
		\bottomrule
	\end{tabular}
\end{table}

\begin{figure}[b!]
	\centering
	\includegraphics[width=8.5cm]{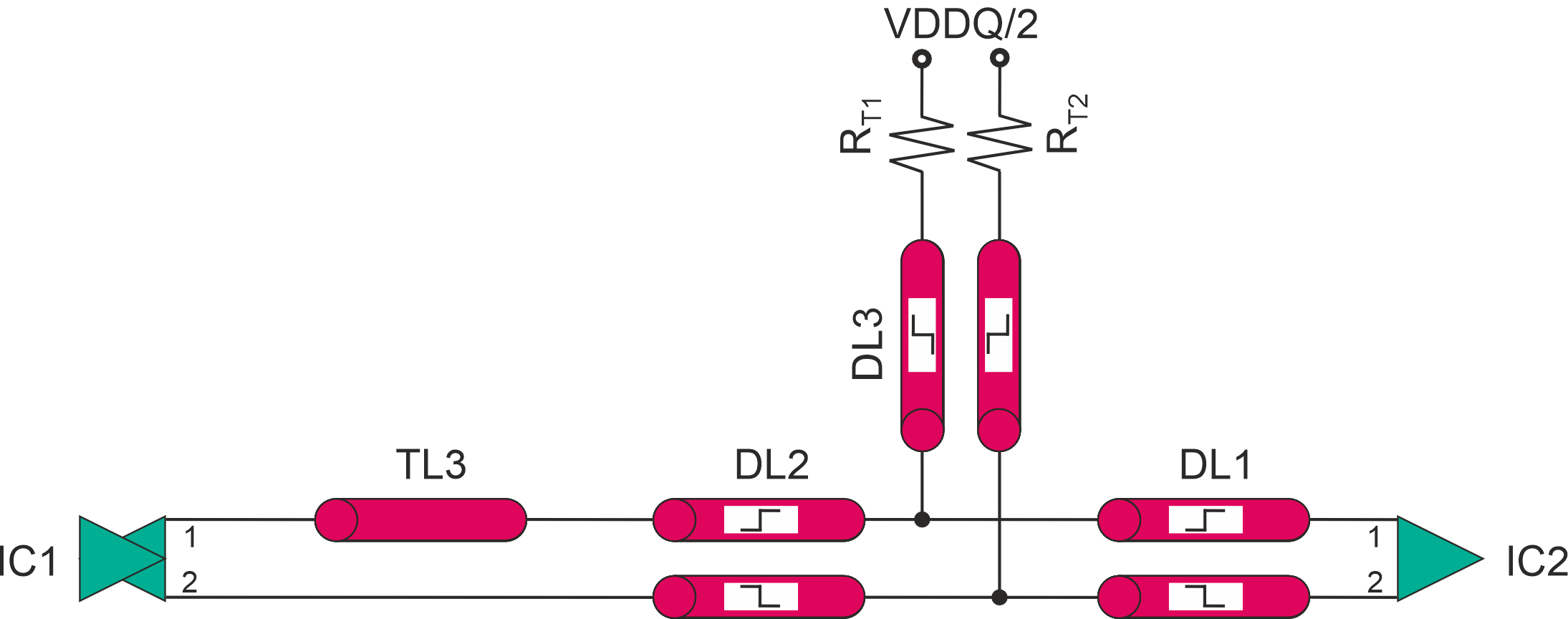}
	\caption{DDR3 DQS interconnect \cite{Withoeft2024}.}
	\label{fig_ddr3_dqs_interconnect}
\end{figure}

Finally, to demonstrate that the framework extends beyond equalization tuning to constrained SI optimization tasks, a DDR3 differential data strobe (DQS) interconnect, shown in Fig.~\ref{fig_ddr3_dqs_interconnect}, was analyzed \cite{Withoeft2024}. In this scenario, the objective was to optimize the microstrip trace geometries and termination to maximize EH and EW while satisfying intra-pair skew constraints. In this setup, the trace width $W$, differential pair spacing $S$, and the termination resistor $R_T$ are the tunable design parameters $T$ to control the differential impedance, coupling, and reflection characteristics. The transmission line routing lengths ($DL_1$, $DL_2$, $DL_3$, $TL_3$) and the dielectric constant $\epsilon_r$ defined the fixed physical layout context $C$. Additionally, to amortize the optimization over varying design specifications, the maximum allowable intra-pair skew threshold was also treated as an explicit dynamically sampled input context $C$ to the framework.

The interconnect consisted of a differential stub series terminated logic (DSSTL) point-to-point structure connecting Tx and Rx via differential transmission lines and a termination branch to half of VDDQ, as typical for DDR3 DQS signals. As shown in Fig.~\ref{fig_ddr3_dqs_interconnect}, the single-ended transmission line $TL_3$ was added to one leg of the differential pair to artificially create a length imbalance, which results in increased intra-pair skew and decreased common-mode rejection.

Data was generated using simulations in the Zuken eCADSTAR Signal Integrity software \cite{eCADSTAR}. The buffer characteristics were described using IBIS models featuring an Intel Arria V for the Tx and a Micron MT41J256M4DA for the Rx, which operate at an I/O clock frequency of 800~MHz (Speedgrade 1600) at 1.5~V. For the eye diagram simulations, a PRBS-10 stimulus was utilized with a 5\% uniformly distributed jitter relative to the UI. A total of 20,000 simulations were generated based on LHS within the parameter ranges specified in Table~\ref{tab_params_ddr3}, which were later divided into an 80-10-10\% train-validation-test split. The trace thickness, dielectric thickness, and loss tangent were kept constant throughout the simulations at 35~µm, 200~µm, and 0.007, respectively.

\section{Amortized Neural Optimization Methodology}

\subsection{Problem Formulation}

For pre-layout SI-DSE, the objective is to rapidly obtain design parameters that comply with specific requirements and optimize the SI metric for a given channel.

Let the fixed interconnect and PCB context parameters be denoted by the vector $C \in \mathbb{R}^{N_C}$, and the tunable design parameters by the vector $T \in \mathbb{R}^{N_T}$. The non-linear physical system is usually evaluated via computational electromagnetic or transient circuit simulations and can be represented as a function $f$
\begin{equation}
	Y = f(C, T),
\end{equation}
where $Y \in \mathbb{R}^{N_Y}$ represents the extracted SI metrics such as EH, EW, inner eye contour, or intra-pair skew. 

Traditional search-based optimization aims to find the optimal settings $T^*$ that minimize a specific performance objective $J(Y)$ within the valid parameter space, denoted by $\mathcal{T}$:
\begin{equation}
	T^* = \arg\min_{T \in \mathcal{T}} J(f(C, T)).
\end{equation}

Because $f(C, T)$ is extracted from computationally expensive simulations, wrapping this system in an iterative search algorithm creates a computational bottleneck, which makes real-time large-scale DSE computationally costly.

\begin{figure*}[t!]
	\centering
	\includegraphics[width=\textwidth]{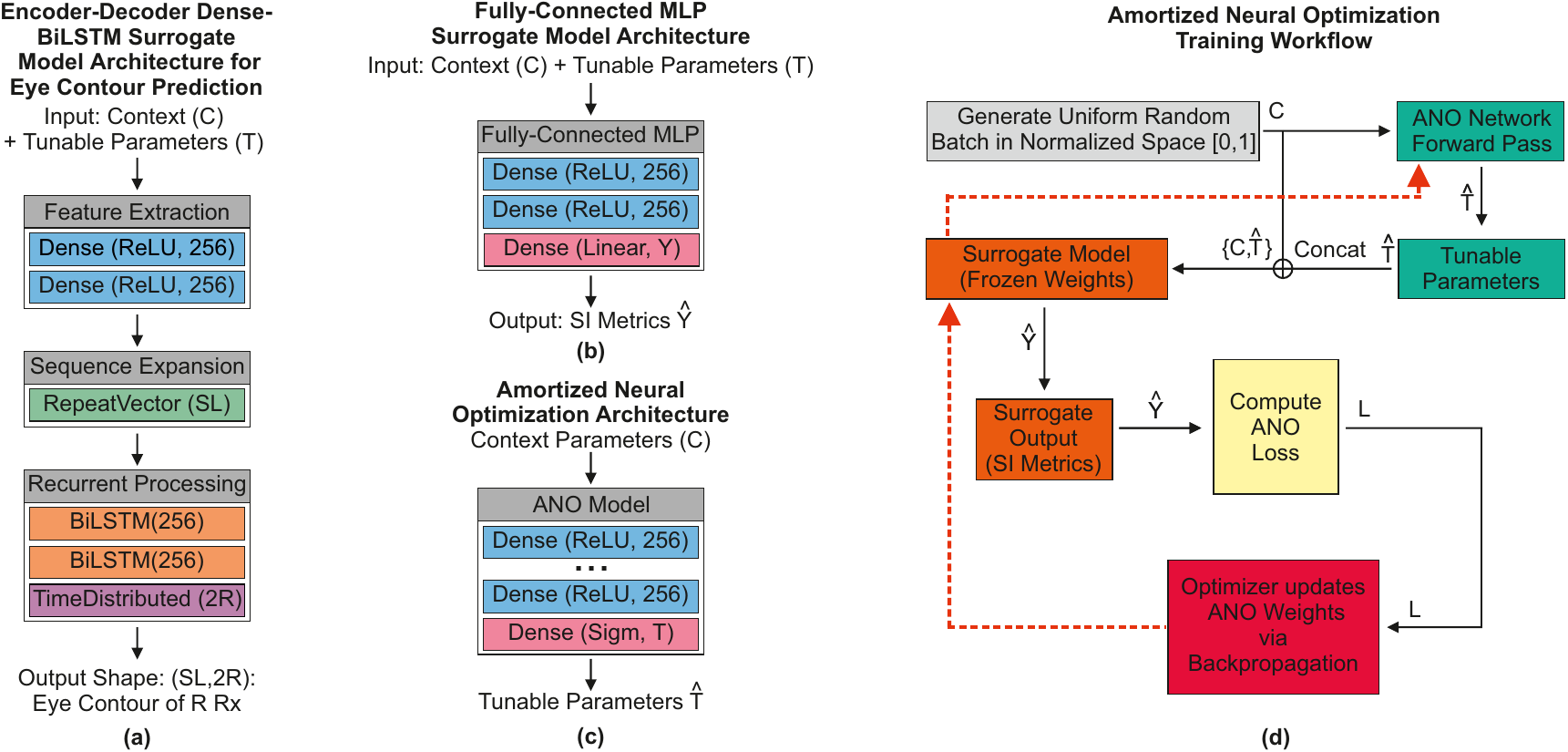}
	\caption{The proposed generalized amortized neural optimization (ANO) framework. The differentiable surrogate models utilizing either an encoder-decoder Dense-BiLSTM architecture (a) specifically for sequential SI metrics such as inner eye contours or a fully-connected MLP architecture (b). The ANO policy network (c) predicts the optimal tunable parameters $\hat{T}$ directly from the given channel context $C$. In the gradient-based training workflow (d), the ANO loss is calculated from the surrogate's predictions and the resulting gradients are backpropagated through the frozen surrogate to update the ANO policy network.}
	\label{fig_ano}
\end{figure*}

\subsection{Differentiable Surrogate Models}

To overcome the computational inefficiencies of simulation, the physical system $f(C, T)$ is approximated by an ML surrogate model $S(C, T; \theta_S)$ parameterized by its learnable parameters $\theta_S$. Fully differentiable neural networks are the natural choice over other partially differentiable regression techniques, especially because the subsequent optimization policy requires continuous gradient flow.

Depending on the nature of the target SI metric $Y$, different neural network architectures can be applied. For predicting sequential data, such as the timesteps of an inner eye contour, an encoder-decoder architecture is utilized. As illustrated in Fig.~\ref{fig_ano} (a), this architecture uses fully-connected layers to transform design parameters into a latent representation. This is followed by sequence expansion and recurrent processing using bidirectional long short-term memory (BiLSTM) layers to capture the temporal structure of the inner eye contour at the output. For scalar metrics, such as EH or intra-pair skew values, a standard multi-layer perceptron (MLP) with only fully-connected layers is utilized, as shown in Fig.~\ref{fig_ano} (b).

Prior to training these surrogate models, all inputs, including $C$ and $T$, were normalized to $[0, 1]$. This ensures training stability across the design parameters, which comprise multiple orders of magnitude. Moreover, it also establishes a uniform space for the optimization phase.

\subsection{Amortized Neural Optimization Architecture}

Black-box optimization algorithms search for the optimal parameters $T^*$ iteratively during the inference phase. In contrast to this iterative optimization, the ANO framework takes a completely different route by training a second neural network. This ANO policy network is denoted as $\Pi(C; \theta_\Pi)$. 

As shown in Fig.~\ref{fig_ano} (c), the ANO policy network only receives the context parameters $C$ as input. Through its hidden fully-connected dense layers, it directly outputs the predicted optimal tunable settings $\hat{T}$. The forward pass of this network maps the channel context to the near-optimal configuration:
\begin{equation}
	\hat{T} = \Pi(C; \theta_\Pi).
\end{equation}

As the underlying surrogate model operates in the normalized $[0, 1]$ space, the final layer of the ANO policy network utilizes a sigmoid activation function. This bounds the ANO policy network's predictions to the parameter ranges that were defined during the data generation phase to make sure that all outputs satisfy the valid design space $\mathcal{T}$.

This dual-network architecture eliminates the need for an iterative optimization algorithm during DSE. Once trained, the policy network provides a near-instantaneous and deterministic prediction of the optimal tunable parameters for any given channel configuration.

\subsection{Amortized Policy Training via Surrogate Gradients}

The core of the ANO methodology lies in the approach used to train the policy network $\Pi$ without requiring an expert dataset of optimal parameter configurations. This is achieved by utilizing the gradient of the fully differentiable surrogate model.

The training workflow is shown in Fig.~\ref{fig_ano} (d). First, uniform random context parameters $C$ are generated per batch. The ANO network performs a forward pass to predict the tunable parameters $\hat{T}$. These predictions and the context parameters $C$ are concatenated and evaluated using the surrogate model. 

During this phase, the weights of the surrogate model $\theta_S$ are frozen, and it then predicts the SI performance metrics, denoted as $\hat{Y}$. A custom loss function $\mathcal{L}$ is then computed based on these predictions, for example, to maximize the target SI performance metrics while penalizing physical constraint violations or performance degradation relative to a baseline configuration.

Because the surrogate is fully differentiable, the gradients of this loss are backpropagated through the frozen surrogate model and directly into the ANO policy network. The optimizer updates only the ANO weights $\theta_\Pi$ to adjust the global policy to output parameter settings that satisfy all physical constraints and at the same time maximize the target SI metrics. Through this process, the computational cost is amortized into the neural network weights. While this requires an upfront offline investment to generate training data to train the surrogate models and then based on that the ANO policy network, it results in a globally reusable asset for the entire design space, which shifts the iterative optimization complexity completely to the offline training phase.

\subsection{Benchmarking Methodology}

To evaluate the proposed ANO framework, its performance is benchmarked against a comprehensive suite of optimization algorithms. To ensure a fair comparison, all baseline algorithms were combined with the same ML surrogate models and using the same objective. The evaluated baselines represent different optimization methodologies:

\begin{itemize}
	\item \textbf{Gradient-Based:} Gradient descent (GD) also directly exploits the continuous differentiability of the neural surrogate based on the objective. GD was realized using the Adam optimizer with a learning rate of 0.05 for a maximum of 200 iterations to ensure convergence while also keeping runtime feasible as each iteration utilizes a computationally more intensive backpropagation pass in addition to the forward pass.
	\item \textbf{Evolutionary/Swarm:} These gradient-free, population-based methods start with a set of candidate solutions and improve these in an iterative manner inspired by biological processes. The evaluated algorithms include the genetic algorithm (GA) \cite{Goldberg1989}, differential evolution (DE) \cite{Storn1997}, and particle swarm optimization (PSO) \cite{Kennedy1995}, all configured with a population size of 100 and executed for 50 generations. Furthermore, the covariance matrix adaptation evolution strategy (CMA-ES) \cite{Hansen2001} was evaluated using an initial step size $\sigma$ of 0.3 with a computational budget of 5,000 surrogate evaluations.
	\item \textbf{Probabilistic:} Bayesian optimization (BO) \cite{Mockus1974} constructs a probabilistic Gaussian process surrogate of the objective function and utilizes an acquisition function to guide the search. BO was executed for a total of 200 optimization steps using expected improvement as the acquisition function. The BO budget had to be limited to 200 steps due to the $\mathcal{O}(N^3)$ cubic scaling complexity of the underlying Gaussian process, making larger evaluation limits computationally prohibitive.
	\item \textbf{Heuristic/Search:} The Nelder-Mead (NM) simplex method \cite{Nelder1965} is based on the geometric transformation of a simplex to navigate the search space, while random search (RS) acts as a baseline stochastic sampler. Both methods were limited to a maximum of 5,000 surrogate evaluations.
\end{itemize}

\subsection{Implementation Details}

For the differentiable surrogate models, training was realized using the Adam optimizer with an initial learning rate of $10^{-3}$ and a batch size of 32. The models were trained to minimize the mean squared error (MSE) loss with a maximum of 1,000 epochs. To ensure convergence and prevent overfitting, a learning rate schedule and an early stopping mechanism with a patience of 50 epochs were utilized.

The ANO policy networks were subsequently trained using the frozen surrogates. Because the policy networks had to evaluate randomized channel contexts simultaneously, they utilized a significantly larger batch size of 2,048 and a smaller Adam learning rate of $10^{-4}$. Specific epoch counts and customized loss formulations were adapted per experiment based on the nature and complexity of the design scenario.

While the trained ANO framework achieves near-instantaneous inference for DSE, it relies on an upfront offline computational investment to simulate the datasets and train the surrogate model and the ANO policy network. For transparency and to quantify this one-time investment, the specific runtimes are reported based on the following workstation hardware equipped with an AMD Ryzen 9 7900 CPU used for data generation, an NVIDIA RTX 4080 Super GPU used for model training, and 64 GB of DDR5 RAM. The DDR5 topology required approximately 136.8~hours for data generation, 51.6~minutes for surrogate training, and 121.5~minutes for ANO policy training. The SerDes link required 48.3~hours, 50.2~minutes, and 28~minutes for the same respective stages. Finally, the DDR3 DQS interconnect required 22.8~hours, 3.3~minutes, and 6.4~minutes. Because these runtimes represent a one-time investment to analyze a high-dimensional channel context that covers a large range of interconnect and technology parameters, this investment results in a fast recovery during large-scale pre-layout DSE.

\section{DDR5 DFE Equalization}

\begin{figure}[t!]
	\centering
	\includegraphics[width=8.5cm]{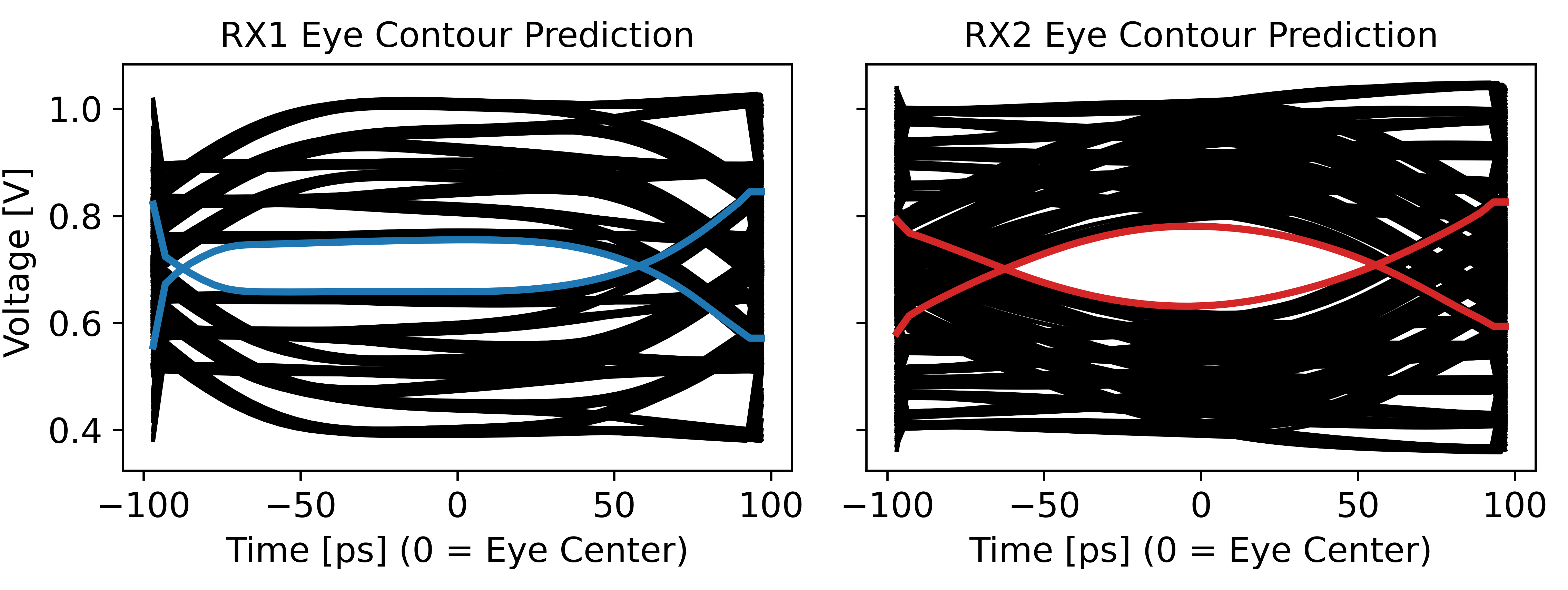}
	\caption{DDR5 encoder-decoder surrogate model prediction example on the test dataset.}
	\label{fig_eye_contour_ddr5}
\end{figure}

The first evaluation of ANO focuses on optimizing the DFE settings $T$ of the two receivers in the DDR5 daisy chain interconnect, shown in Fig.~\ref{fig_ddr5_interconnect}, based on the given PCB context parameters $C$ \cite{Withoeft2026}. The first requirement for the ANO framework is an accurate and differentiable approximation of the DDR5 interconnect. Table~\ref{tab_regression_metrics_ddr5} confirms that the Dense-BiLSTM encoder-decoder surrogate, see Fig.~\ref{fig_ano} (a), achieves accurate regression metrics on the unseen test dataset with $R^2$-scores above 96\% and a sample-wise normalized root mean squared error (NRMSE) below 7\% across both receivers. This sample-wise metric scales the absolute error not by the global minimum and maximum, but rather by the maximum vertical swing of the inner eye contour to ensure meaningful errors relative to the actual eye opening. Fig.~\ref{fig_eye_contour_ddr5} validates this visually and demonstrates that the surrogate predicts inner eye contours which match the ground-truth simulations well. This accurate and differentiable backbone is what enables the ANO framework to perform gradient-based policy learning.

For this dual-receiver DFE optimization task, the ANO policy network was designed with a shared feature-extraction trunk and two independent heads, which output the 4-tap settings for Rx1 and Rx2. The ANO network was trained for 50 epochs. To compute EH during training, it was extracted from the predicted 100-dimensional inner eye contour defined as the minimum vertical distance between the upper and lower contour within a sampling window at the center of the UI. The objective was to maximize the post-DFE EH improvement, defined as $g = \mathrm{EH}_{\mathrm{DFE}} - \mathrm{EH}_{\mathrm{DFE\;off}}$ at each receiver. To achieve this, the network was trained using a custom loss function $L(g)$:
\begin{equation}
	\label{eq_loss}
	L(g) = \underbrace{\lambda_1 \left[ \max(0, -g) \right]^2}_{\text{Quadratic Penalty}} - \underbrace{\lambda_2 \ln \left( 1 + \max(0, g) \cdot \alpha \right)}_{\text{Logarithmic Reward}},
\end{equation}
where the hyperparameters were set to $\lambda_1 = 1000$, $\lambda_2 = 1$, and $\alpha = 100$. This is an asymmetric optimization policy. The first term leads to a quadratic penalty to prevent the network from choosing DFE configurations that degrade the EH compared to the DFE-off setting. The second term provides a logarithmic reward for EH improvements with a diminishing gradient magnitude of the logarithm to prevent the optimizer from overfitting to simple channel instances and to ensure convergence for the entire design space.

To prevent large gradients from the receiver with heavier signal degradation, which would lead to domination of the shared network weights, a multi-task gradient balancing technique was utilized. During backpropagation, the gradient vectors for $L(g_{Rx1})$ and $L(g_{Rx2})$ were computed independently and divided by their respective magnitudes. This normalized both gradient updates to a unit length before they were summed to ensure that both receiver heads contributed equally to the optimization of the shared trunk.

\begin{table}[t!]
	\centering
	\caption{Encoder-decoder surrogate model regression accuracy metrics for DDR5 eye contour prediction on the test dataset, as established in \cite{Withoeft2026}}
	\label{tab_regression_metrics_ddr5}
	\begin{tabular}{lccc}
		\toprule
		\textbf{Receiver} & \textbf{RMSE [mV]} & \textbf{NRMSE [\%]} & $\mathbf{R^2}$ [\textbf{\%}] \\
		\midrule
		Rx1 & 27.2 & 6.46 & 96.11 \\
		Rx2 & 23.6 & 6.98 & 96.55 \\
		\bottomrule
	\end{tabular}
\end{table}

\begin{figure}[t!]
	\centering
	\includegraphics[width=8.5cm]{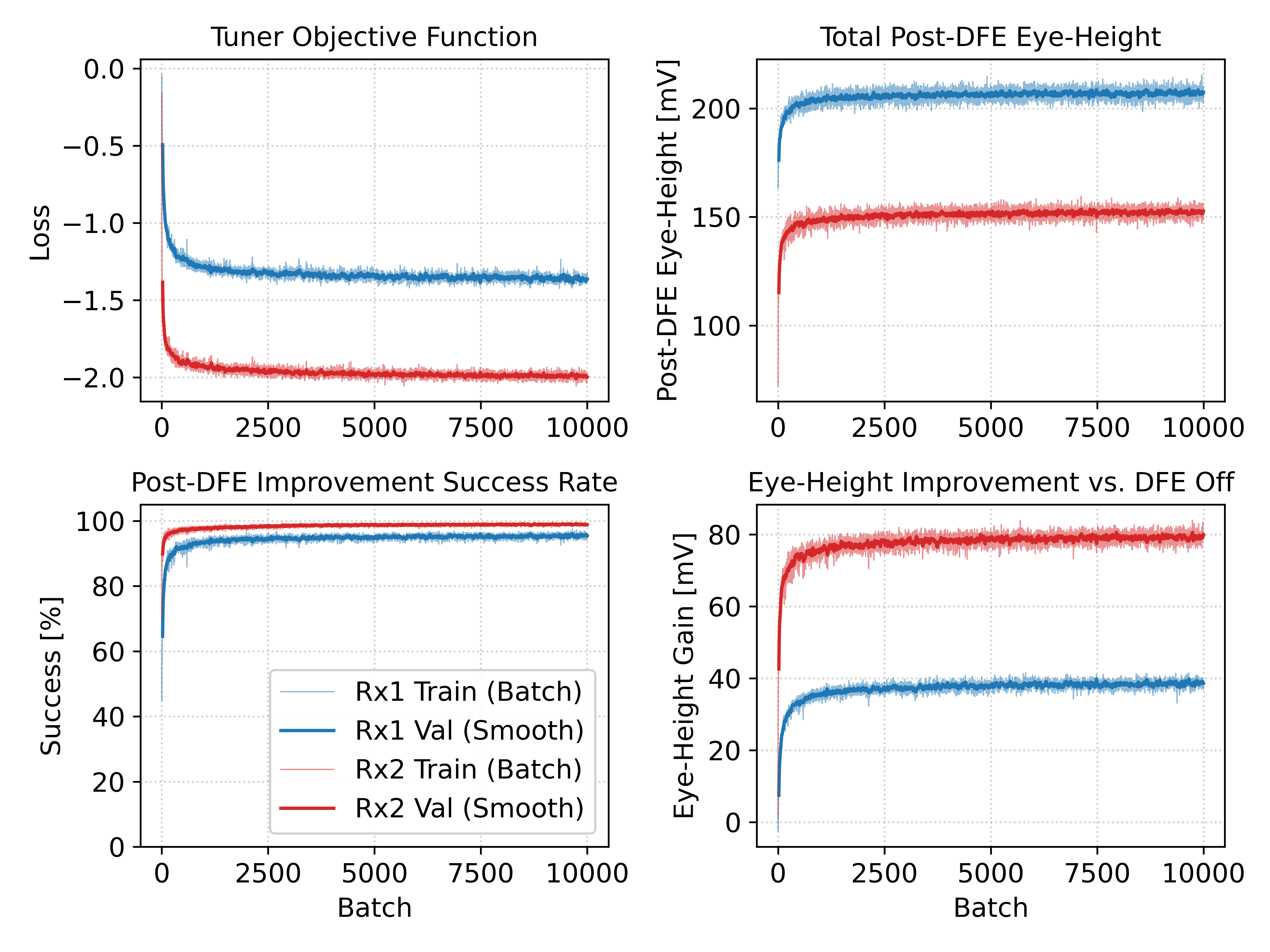}
	\caption{DDR5 ANO training curves showing loss, post-DFE EH, success rates, and ANO-based EH improvement for both Rx1 and Rx2.}
	\label{fig_training_ddr5}
\end{figure}

The training process of the ANO policy network is shown in Fig.~\ref{fig_training_ddr5}. The loss function converged quickly and then stabilized within the first 2,500 training batches. The network achieved a final improvement success rate of over 95\% at both receivers. Because Rx2 is located farther from the Tx in the daisy chain and is therefore more heavily impaired by ISI and signal attenuation, the ANO network learned to apply more aggressive equalization, which resulted in large post-DFE EH improvements of roughly 80~mV. There was less channel degradation at Rx1, which resulted in smaller improvements of approximately 40~mV.

\begin{table}[b!]
	\centering
	\caption{DDR5 benchmark comparison (means) of optimization methods averaged over 100 randomized channel contexts using CPU and GPU-accelerated surrogate/ANO evaluations}
	\label{tab_benchmark_comparison_ddr5}
	\begin{tabular}{lccccc}
		\toprule
		\textbf{Method} & Success & $g_{Rx1}$ & $g_{Rx2}$ & $t_{CPU}$ & $t_{GPU}$ \\
		\midrule
		ANO				& 96\%  & 41.4 mV & 82.3 mV & 0.001 s & 0.004 s \\
		GD				& 89\%  & 47.2 mV & 73.2 mV & 7.5 s   & 7.8 s \\
		GA				& 100\% & 51.4 mV & 84.5 mV & 17.2 s  & 2.7 s \\
		DE				& 100\% & 51.5 mV & 84.4 mV & 17.1 s  & 2.8 s \\
		PSO				& 100\% & 50.9 mV & 84.3 mV & 17.6 s  & 2.8 s \\
		CMA-ES			& 92\%  & 52.9 mV & 74.7 mV & 53.7 s  & 10.3 s \\
		BO				& 96\%  & 49.6 mV & 73.3 mV & 136.4 s & 163.4 s \\
		NM				& 32\%  & 4.0 mV  & 15.1 mV & 63.8 s  & 21.7 s \\
		RS				& 83\%  & 27.9 mV & 56.4 mV & 16.1 s  & 0.160 s \\
		\bottomrule
	\end{tabular}
	
	\vspace{2pt}
	\parbox{\linewidth}{\centering\scriptsize\textit{Hardware:} AMD Ryzen 9 7900, NVIDIA RTX 4080 Super, 64 GB DDR5 RAM}
\end{table}

To assess the accuracy and efficiency of the ANO framework, it was evaluated against the suite of optimization methods. The results are summarized in Table~\ref{tab_benchmark_comparison_ddr5} and indicate a clear difference in computational efficiency. Population-based evolutionary and swarm algorithms, such as GA, DE, and PSO, achieve near-perfect success rates and slightly higher absolute EH improvements. However, this marginal performance edge comes at a computational cost of over 17~seconds on the CPU and 2.7~seconds on the GPU per channel due to the thousands of sequential surrogate evaluations required per optimization run. CMA-ES only performs slightly worse at Rx2 and slightly better at Rx1 compared to the other evolutionary algorithms, and is even more computationally intensive at 53~seconds on the CPU. BO similarly achieves a high success rate of 96\% and solid EH improvements, but it is even more bottlenecked computationally due to the cubic scaling complexity of the underlying Gaussian process, which results in an impractical execution time of over 136~seconds on the CPU per channel.

The ANO framework predicts the near-optimal DFE taps using one single forward pass, which reduces the inference time to just 1~millisecond on the CPU and 4~milliseconds on the GPU, because there is overhead on the GPU due to data transfer for a single evaluation. This is a speedup of three to four orders of magnitude compared to evolutionary and swarm methods. GD also makes use of the surrogate’s differentiability, but is still iterative and computationally expensive due to repeated forward and backpropagation steps for each new channel. GD takes 7.5~seconds for the optimization and results in slightly worse performance compared to ANO on average. Simpler methods such as NM and RS achieve subpar optimization results in these complex SI design scenarios. NM only achieves a 32\% success rate with small EH improvements, while RS also leads to sub-optimal EH improvements.

\begin{figure}[b!]
	\centering
	\includegraphics[width=8.5cm]{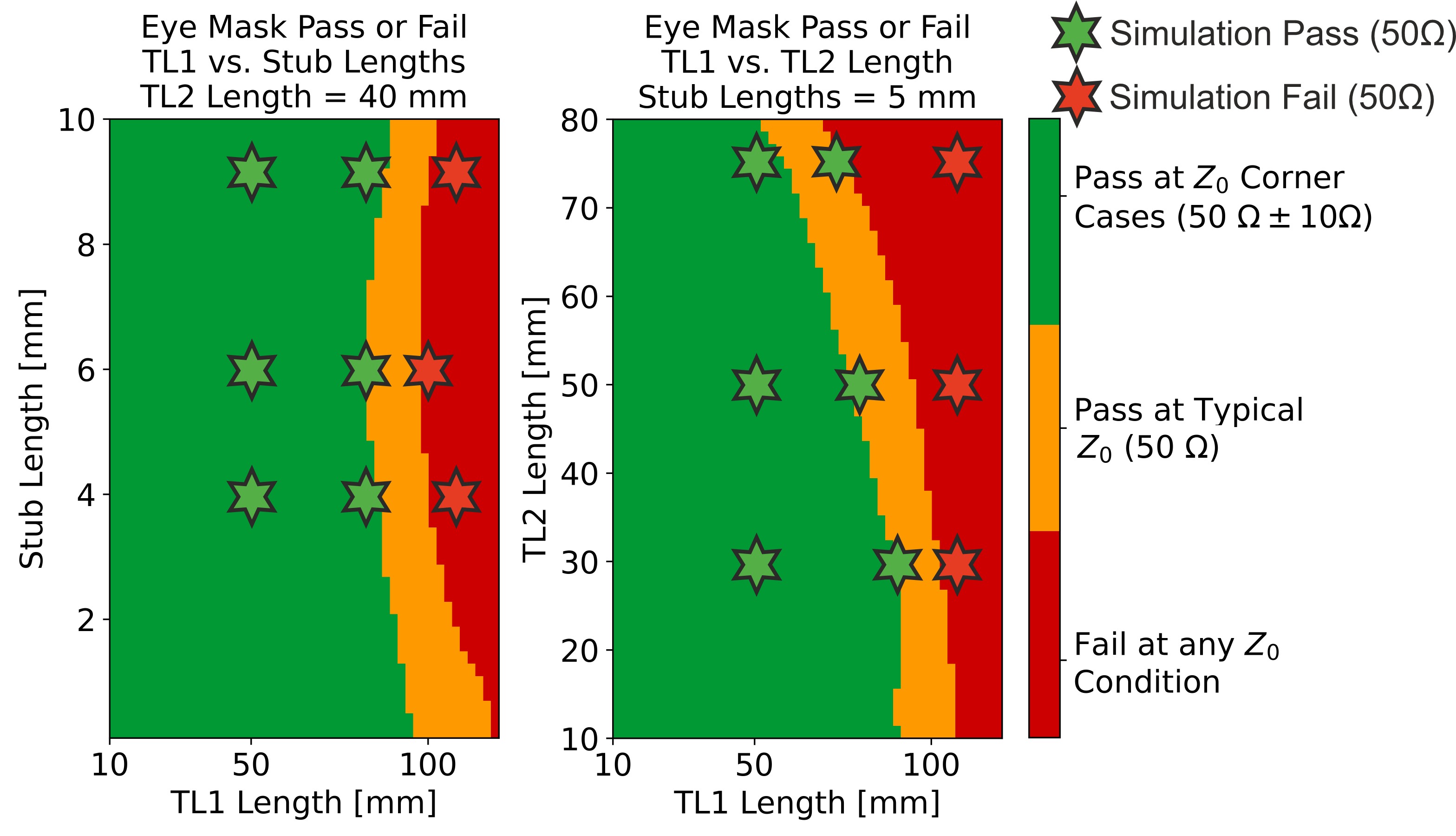}
	\caption{SI design space exploration for the DDR5 interconnect using the optimized DFE taps. Adapted from \cite{Withoeft2026} with explicit simulation verification denoted by stars to validate the ANO and surrogate boundary predictions.}
	\label{fig_dse_ddr5}
\end{figure}

The true practical value of this near-instantaneous inference is demonstrated based on the large-scale SI-DSE shown in Fig.~\ref{fig_dse_ddr5} \cite{Withoeft2026}. The green regions show feasible design configurations where the ANO-predicted DFE settings satisfy the rectangular-shaped eye mask with 70~mV height and 0.4~UI width for all of the corner cases $50~\Omega \pm 10~\Omega$. Orange regions show design areas where compliance is only achievable under typical $50~\Omega$ conditions, while for the red infeasible regions no equalization strategy can rescue the signal. The compliance predictions generated by the ANO framework using the ANO policy DFE settings and the surrogate model for contour prediction were additionally validated using simulations denoted by stars in Fig.~\ref{fig_dse_ddr5}, which confirmed the accuracy of the predicted feasible design space.

This large-scale sweep of a $50 \times 50$ transmission line length grid across three global $Z_0$ corner cases leads to a total of 7,500 unique channel configurations and exposes the scaling limitations of the other optimization algorithms. Based on the single-run CPU timings of Table~\ref{tab_benchmark_comparison_ddr5}, it would take approximately 15~hours using GD, 35~hours using standard evolutionary and swarm methods (GA, DE, PSO), over 4~days with CMA-ES, and nearly 12~days using BO to sequentially optimize this grid. Even on a GPU, multiple hours of runtime are to be expected. Even though multi-threaded execution can proportionally reduce the runtime for these iterative algorithms, their aggregate computational complexity remains large. In contrast to these iterative methods, the ANO policy network predicts the entire 7,500-point sweep as a single batched input matrix in just 30~milliseconds on the CPU, and under 2~milliseconds when GPU-accelerated. This transforms an otherwise infeasible or highly resource-intensive optimization task into a near-instantaneous pre-layout SI-DSE tool.

\section{SerDes Co-Equalization}

\begin{table}[b!]
	\centering
	\caption{Regression accuracy metrics for MLP vs. Dense-BiLSTM surrogate models for SerDes eye contour prediction on the test dataset}
	\label{tab_regression_metrics_serdes}
	\begin{tabular}{lccc}
		\toprule
		\textbf{Model} & \textbf{RMSE [mV]} & \textbf{NRMSE [\%]} & $\mathbf{R^2}$ \textbf{[\%]} \\
		\midrule
		MLP & 23.21 & 7.05 & 88.86 \\
		Enc-Dec & 16.62 & 4.39 & 94.35 \\
		\bottomrule
	\end{tabular}
\end{table}

\begin{figure}[b!]
	\centering
	\includegraphics[width=8.5cm]{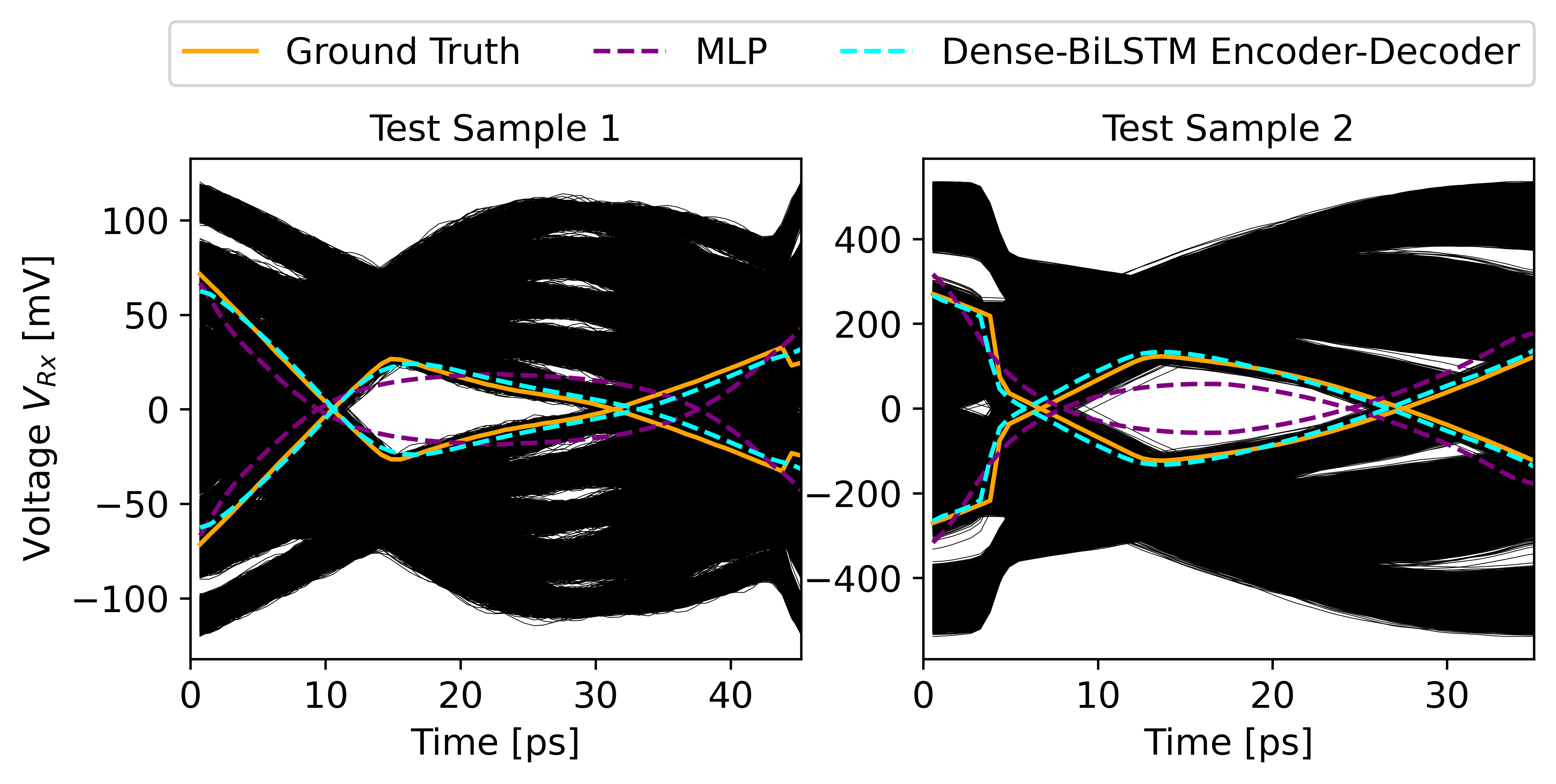}
	\caption{Two SerDes prediction examples of the MLP and encoder-decoder surrogate models on the test dataset.}
	\label{fig_eye_contour_serdes}
\end{figure}

Following the demonstration of Rx-side DFE optimization in the DDR5 topology, the ANO framework was evaluated for complex co-equalization of the SerDes interconnect, which requires the simultaneous optimization of 9 equalization parameters $T$ distributed across both the transmitter with FFE and receiver with CTLE and DFE to counteract frequency-dependent loss and ISI across a 25-dimensional design space context $C$. To train the gradient-based ANO policy, an accurate and fully differentiable surrogate model had to be trained first to approximate the inner eye contour of the SerDes link at the receiver. Two neural network architectures were trained and compared, namely a standard fully-connected MLP, see Fig.~\ref{fig_ano} (b), and a Dense-BiLSTM encoder-decoder, see Fig.~\ref{fig_ano} (a). The prediction performance of both models on the unseen test dataset is summarized in Table~\ref{tab_regression_metrics_serdes}. The MLP predicts a reasonable approximation, but the encoder-decoder architecture is much more accurate throughout all metrics, as it reduces the error significantly and improves the $R^2$-score by more than 5\%.

The underlying reason for this difference in performance becomes clear when inspecting the predicted inner eye contours in comparison to the simulated ground truth, as shown in Fig.~\ref{fig_eye_contour_serdes} for two randomly selected test samples. The MLP predictions shown in dashed purple cannot predict the non-linear temporal dynamics of the eye contour accurately, as they tend to smooth the contour and miss the maximum vertical opening. In contrast, the Dense-BiLSTM encoder-decoder shown in dashed cyan utilizes its recurrent layers to model the sequential dependencies of the eye contour, which allows it to predict the ground-truth contour shown in solid orange and its temporal details more closely. Due to this higher accuracy, the encoder-decoder model was deployed as the differentiable surrogate backbone for the following SerDes ANO policy training.

\begin{figure}[t!]
	\centering
	\includegraphics[width=8.5cm]{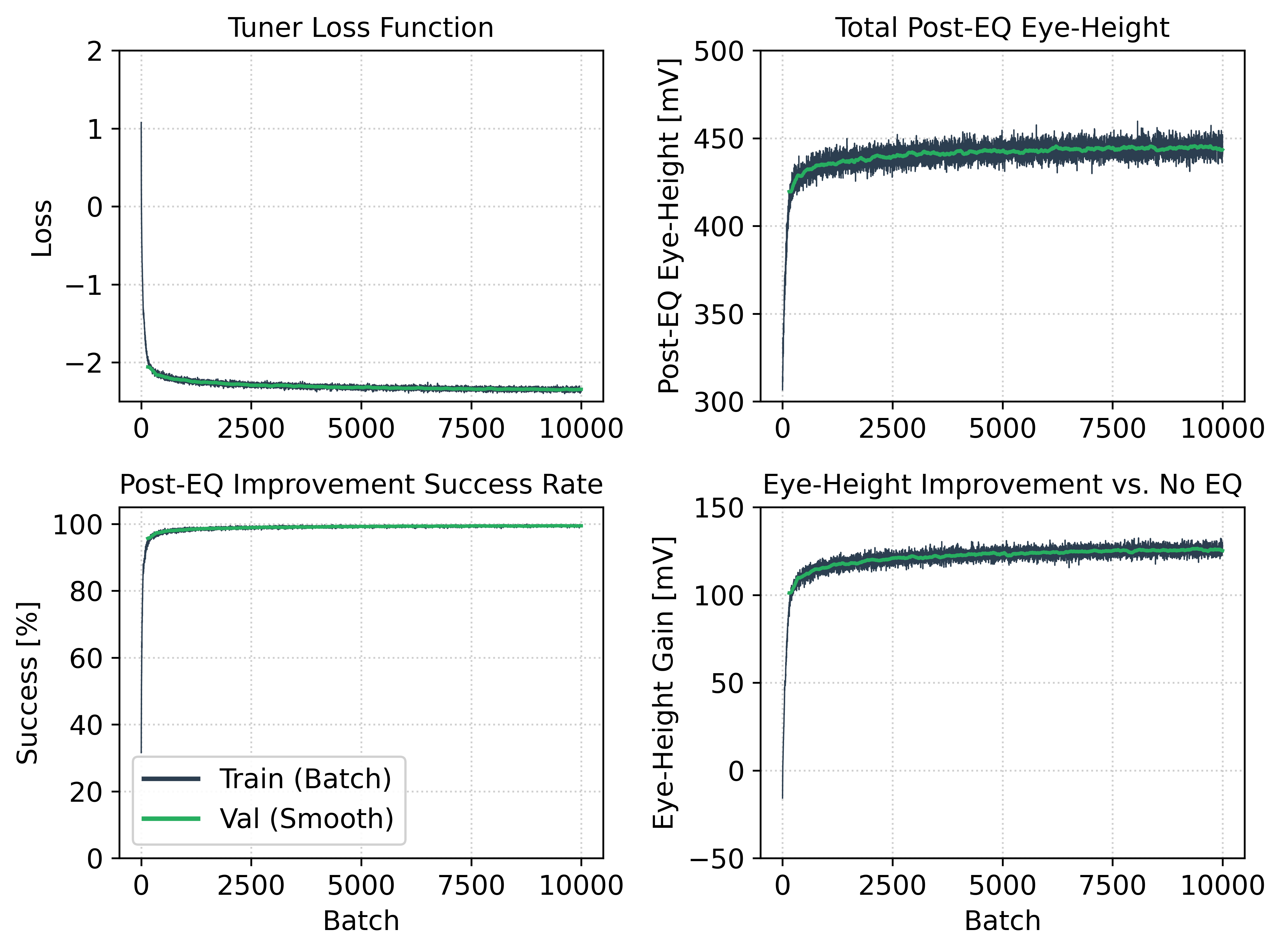}
	\caption{SerDes ANO training curves showing loss, post-EQ EH, success rate, and ANO-based EQ EH improvement vs. unequalized baseline.}
	\label{fig_training_serdes}
\end{figure}

The ANO policy network was trained to co-optimize the 9 equalization parameters, including 2 Tx FFE taps, 3 Rx CTLE parameters, and 4 Rx DFE taps, based on the 25-dimensional channel context parameters. The target of maximizing the EH improvement $g = \mathrm{EH}_{\mathrm{EQ}} - \mathrm{EH}_{\mathrm{EQ\;off}}$ remains the same as for the DDR5 scenario, and therefore the policy network was trained using the same asymmetric loss formulation defined in Eq.~(\ref{eq_loss}). The EH was similarly extracted from the surrogate's 128-dimensional predicted eye contour at the UI center. However, in contrast to the DDR5 scenario, this network utilized a single output head to predict all of the tunable equalization parameters together.

\begin{table}[t!]
	\centering
	\caption{SerDes benchmark comparison (means) of co-equalization optimization methods over 100 randomized channel contexts using CPU and GPU-accelerated surrogate/ANO evaluations}
	\label{tab_benchmark_comparison_serdes}
	\begin{tabular}{lcccc}
		\toprule
		\textbf{Method} & EH & EH Gain & $t_{CPU}$ & $t_{GPU}$ \\
		\midrule
		ANO				& 424.2 mV & 132.5 mV & 0.0006 s & 0.003 s \\
		GD				& 340.6 mV & 48.9 mV  & 8.9 s    & 5.0 s \\
		GA				& 437.8 mV & 146.2 mV & 26.5 s   & 2.1 s \\
		DE				& 439.7 mV & 148.0 mV & 26.3 s   & 2.2 s \\
		PSO				& 439.0 mV & 147.3 mV & 26.9 s   & 2.3 s \\
		CMA-ES			& 421.7 mV & 130.0 mV & 28.3 s   & 5.3 s \\
		BO				& 413.1 mV & 121.4 mV & 114.2 s  & 173.4 s \\
		NM				& 312.6 mV & 20.9 mV  & 23.9 s   & 8.8 s \\
		RS				& 375.5 mV & 83.8 mV  & 20.5 s   & 0.149 s \\
		\bottomrule
	\end{tabular}
	
	\vspace{2pt}
	\parbox{\linewidth}{\centering\scriptsize\textit{Hardware:} AMD Ryzen 9 7900, NVIDIA RTX 4080 Super, 64 GB DDR5 RAM}		
\end{table}

The training process of the SerDes ANO policy is shown in Fig.~\ref{fig_training_serdes}. Within 50 epochs, the network achieved a post-EQ improvement success rate of almost 100\% on the validation set. The ANO policy network learned the synergies between FFE, CTLE, and DFE and improved the total EH to an average of approximately 424~mV and achieved a significant mean EH improvement of over 130~mV compared to the unequalized baseline.

\begin{figure}[t!]
	\centering
	\includegraphics[width=8.5cm]{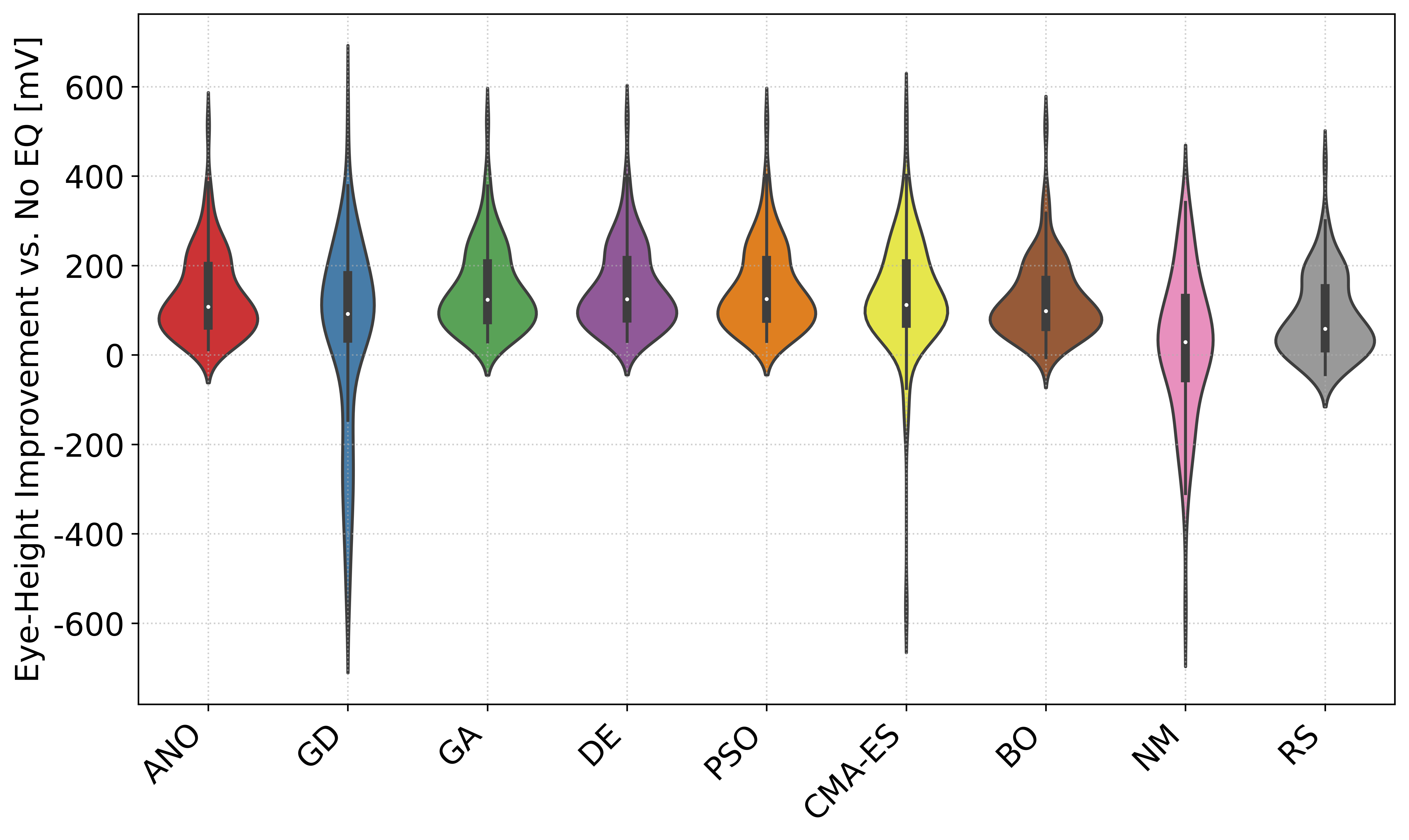}
	\caption{Violin plots statistically visualizing the EH improvement distributions of the benchmark comparison in Table~\ref{tab_benchmark_comparison_serdes}.}
	\label{fig_violin_serdes}
\end{figure}

To evaluate the accuracy and computational efficiency of the ANO policy, its performance was benchmarked against the suite of optimization algorithms based on 100 randomized SerDes channel contexts. The results are summarized in Table~\ref{tab_benchmark_comparison_serdes}, and the EH improvement distributions are visualized using violin plots in Fig.~\ref{fig_violin_serdes}. The findings further confirm the computational efficiency observed in the DDR5 scenario.

\begin{figure*}[b!]
	\centering
	\includegraphics[width=\textwidth]{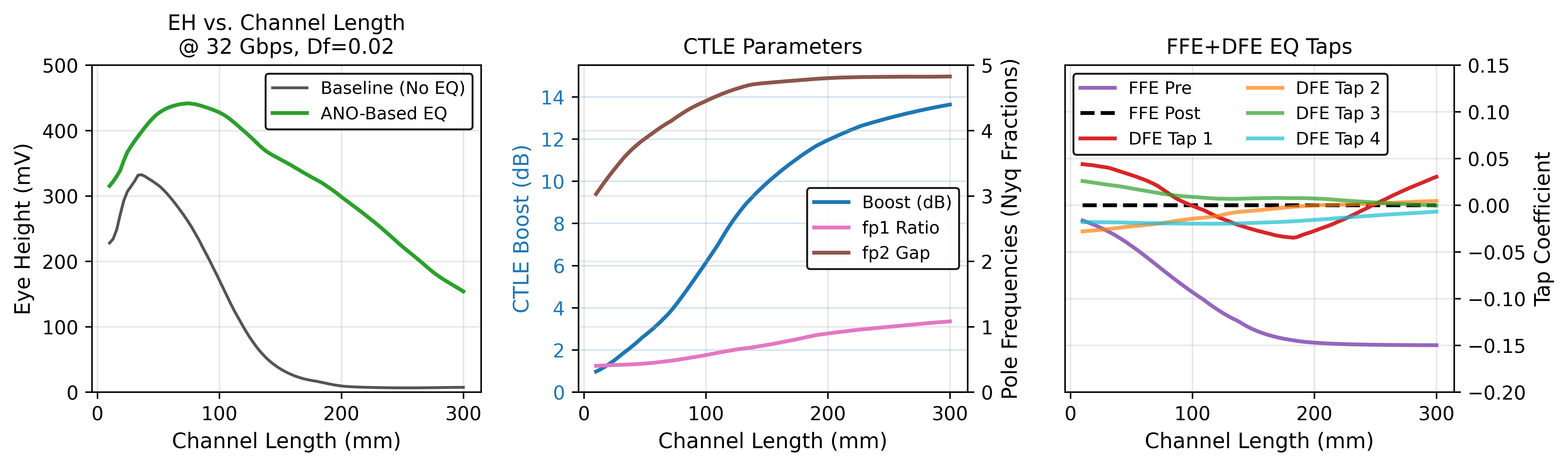}
	\caption{1-dimensional channel length sweep to validate the physical consistency of the 9-dimensional ANO-optimized equalization parameters.}
	\label{fig_1d_sweep_serdes}
\end{figure*}

Population-based evolutionary and swarm algorithms (GA, DE, and PSO) achieve the highest average EH gains at approximately 147~mV. However, the violin plots show that ANO maintains a comparable performance distribution compared to these population-based optimization methods. In contrast, GD and RS show significantly larger variances and more frequent convergence to sub-optimal local minima within the 9-dimensional equalization space. NM degrades heavily in this high-dimensional and complex setting, while BO and CMA-ES are performance-wise close to ANO in this comparison.

Although the evolutionary and swarm approaches achieve roughly 15~mV of additional eye opening on average, this difference reflects the amortization gap between a learned global policy and instance-specific iterative optimization. In many practical design space exploration scenarios, sacrificing approximately 10\% of the maximum achievable improvement represents a beneficial trade-off given the significant reduction in runtime. The population-based iterative methods require more than 26~seconds per run on the CPU due to the 5,000 sequential surrogate evaluations performed during each optimization run. In direct comparison, ANO predicts the equalizer parameters in a single deterministic forward pass that only takes 0.6~milliseconds on a standard CPU and therefore provides a speedup that exceeds four orders of magnitude. This demonstrates that ANO realizes a computationally efficient alternative to iterative black-box search while keeping performance close to these best-performing strategies. This characteristic positions the ANO framework as a strong candidate for real-time Tx/Rx co-equalization optimization scenarios in high-speed SerDes links.

Before deploying the ANO framework for the high-dimensional SerDes design space exploration, its physical consistency had to be verified using a fixed context $C$ at a data rate of 32~Gbps with only a 1-dimensional channel length sweep, as shown in Fig.~\ref{fig_1d_sweep_serdes}. As channel length $L$ increases and high-frequency attenuation becomes more pronounced, the ANO policy performs a smooth and controlled strategy that is physically consistent and interpretable. To compensate for the rising insertion loss, the CTLE boost is progressively ramped up toward its 15~dB limit, while the first and second pole frequencies are increased simultaneously, all of which aim to mitigate the rising high-frequency attenuation. Concurrently, the Tx FFE pre-cursor tap is activated to mitigate pre-cursor ISI, while the policy relies on the Rx DFE taps to cancel the remaining post-cursor ISI. The policy keeps the Tx FFE post-cursor at zero across the entire range, which demonstrates that the model has discovered a power-optimal equalization strategy for this environment. By delegating all post-cursor cancellation to the Rx-side CTLE and the nonlinear idealized 4-tap DFE, the policy avoids the voltage penalty of the Tx FFE post-tap, thereby preserving Tx output amplitude.

\begin{figure*}[t!]
	\centering
	\includegraphics[width=\textwidth]{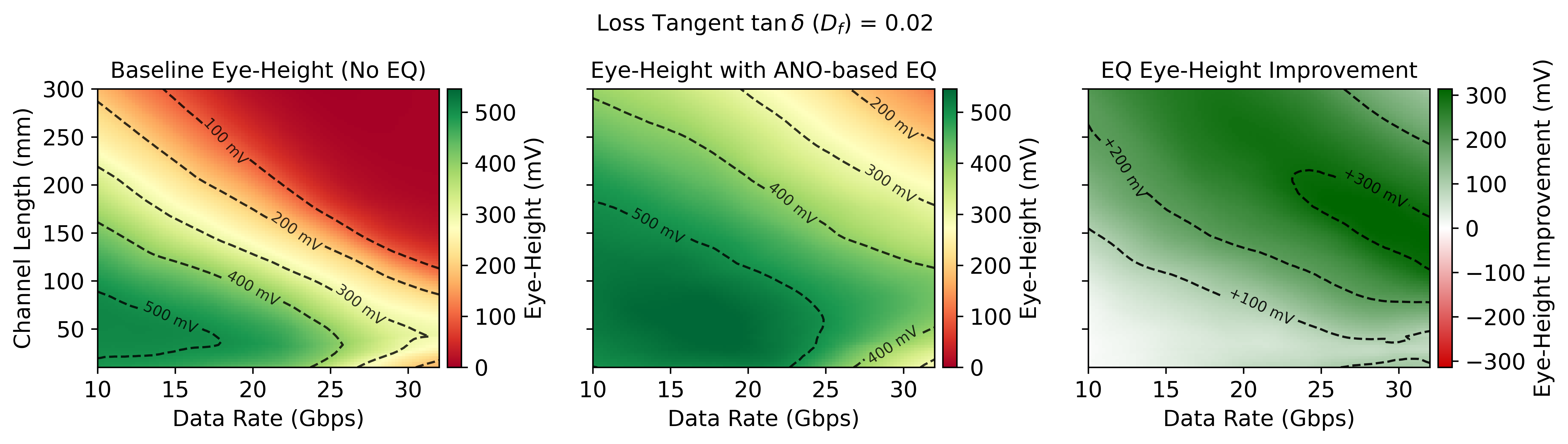}
	\caption{Large-scale design space exploration example using the ANO-based EQ and surrogate predictions of worst-case EH performance over a $100 \times 100$ grid of data rates and channel lengths at 32 manufacturing corner cases and a nominal loss tangent ($D_f=0.02$) for the unequalized baseline (left), ANO-based EQ (center), and ANO-based EQ EH improvement (right).}
	\label{fig_dse_serdes}
\end{figure*}

\begin{figure}[b!]
	\centering
	\includegraphics[width=8.5cm]{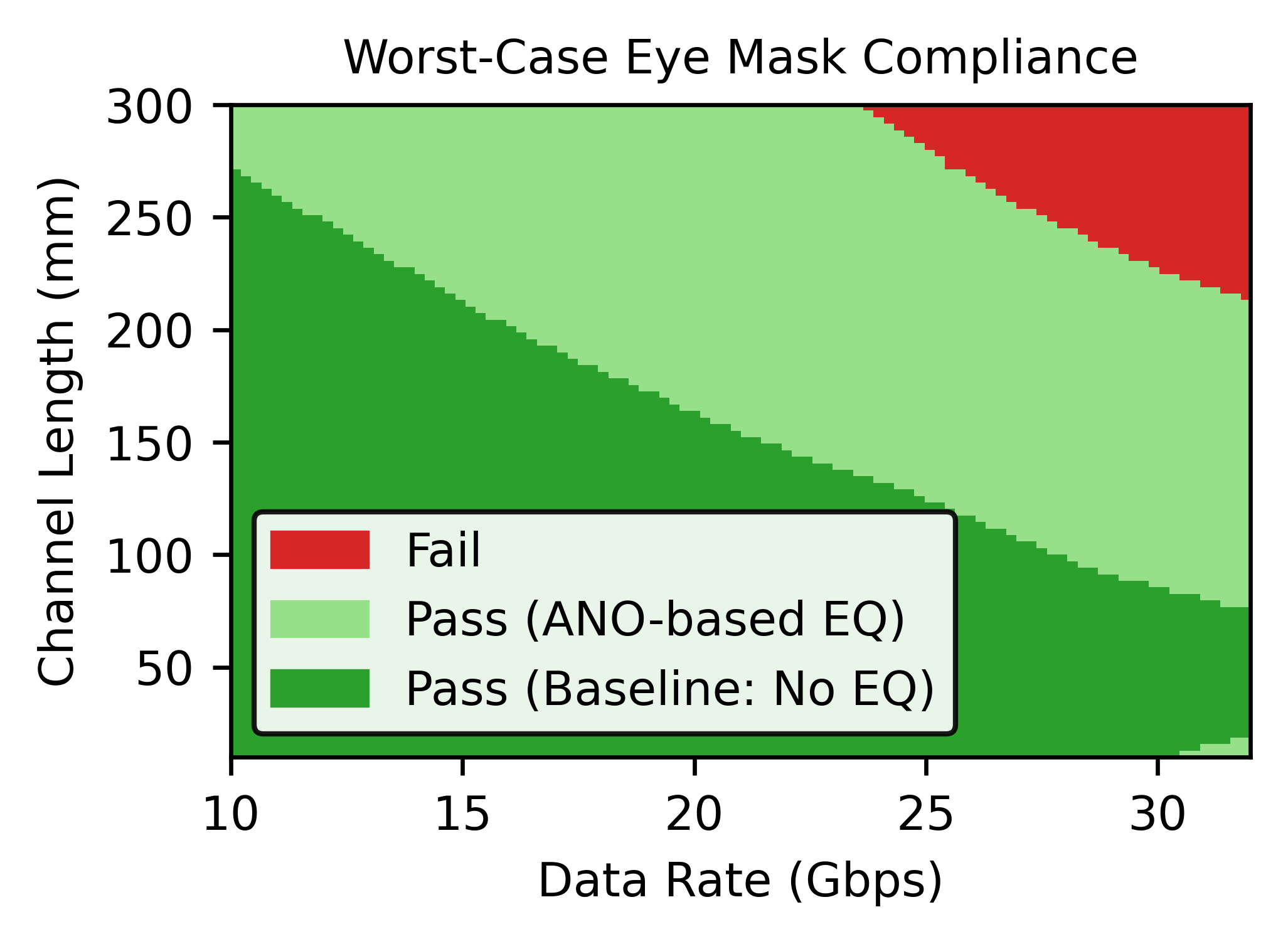}
	\caption{Worst-case eye mask compliance feasibility boundary across all 32 manufacturing corner cases.}
	\label{fig_eye_mask_serdes}
\end{figure}

To demonstrate the potential of the ANO framework for real-world EDA applications, DSE was performed across a dense $100 \times 100$ grid of channel lengths (10 to 300~mm) and data rates (10 to 32~Gbps). To include manufacturing tolerances, each of the 10,000 grid points was evaluated across $2^5=32$ corner cases spanning $\pm5\%$ process tolerances to the nominal values of trace width ($W=100~\mu$m), trace spacing ($S=200~\mu$m), substrate height ($H=250~\mu$m), dielectric constant $\epsilon_r$ ($D_k=4.5$), and loss tangent $\tan\delta$ ($D_f=0.02$), which led to a nominal differential impedance $Z_{\mathrm{diff}}$ of approximately 100~$\Omega$. This corresponds to 32 manufacturing corner evaluations per grid point, resulting in a total of 320,000 channel configurations and ANO optimization runs.

The aggregated worst-case EH performance for these 320,000 evaluations is visualized in Fig.~\ref{fig_dse_serdes}. The unequalized baseline (left) shows eye closure for larger lengths and data rates, while the ANO-based equalization (center) successfully rescues the eye across the majority of the design space. The ANO-based EH improvement (right) shows that the ANO framework provides an additional worst-case eye opening of up to 300~mV in high-loss regions to effectively improve the feasible interconnect length even under manufacturing tolerances.

Additionally, two further $100 \times 100$ grid sweeps incorporating the same $\pm5\%$ manufacturing tolerances with loss tangent vs. channel length and loss tangent vs. data rate have been considered. The results of these experiments and the specific equalization settings for these scenarios and the one in Fig.~\ref{fig_dse_serdes} are shown in the Appendix \ref{cha_appendix}.

To analyze how these EH improvements lead to practical pre-layout DSE decisions, Fig.~\ref{fig_eye_mask_serdes} shows the worst-case eye mask compliance. A design is only classified as a ``Pass'' if the ANO-predicted equalizer settings produce eye contours that satisfy a hexagonal eye mask for all 32 independent manufacturing corners simultaneously. The eye mask has a minimum vertical eye opening of 200~mV peak-to-peak ($\pm 100$~mV) for the central region with a flat top of 0.25~UI and symmetrical slopes that terminate at 0~V at $\pm 0.25$~UI from the UI center. This analysis then distinguishes safe designs even without equalization shown in dark green from those salvaged by ANO-based equalization in light green and unrecoverable failure zones in red.

Finally, this 320,000-point multi-corner sweep provides even stronger evidence for the scaling advantage of the ANO framework. Based on the runtimes of Table~\ref{tab_benchmark_comparison_serdes}, this multi-corner optimization using an iterative optimization algorithm (averaging 26.5~s on a CPU and 2.2~s on a GPU) would require approximately 98~days of continuous computation on a CPU ($320,000 \times 26.5$~s) and 8~days on a GPU ($320,000 \times 2.2$~s). In contrast, the ANO policy processes the entire 320,000-point batch vectorized as a single matrix operation. On a CPU, this large sweep optimization run is executed in just 280~milliseconds. With GPU acceleration, the inference time for all 320,000 configurations reduces to only 9.3~milliseconds. The computational cost is therefore reduced from days to milliseconds. The ANO framework transforms computationally expensive multi-corner pre-layout sweeps into real-time feedback for high-speed DSE and worst-case analysis.

\section{DDR3 DQS Eye Optimization under Intra-Pair Skew Constraints}

\begin{figure}[t!]
	\centering
	\includegraphics[width=8.5cm]{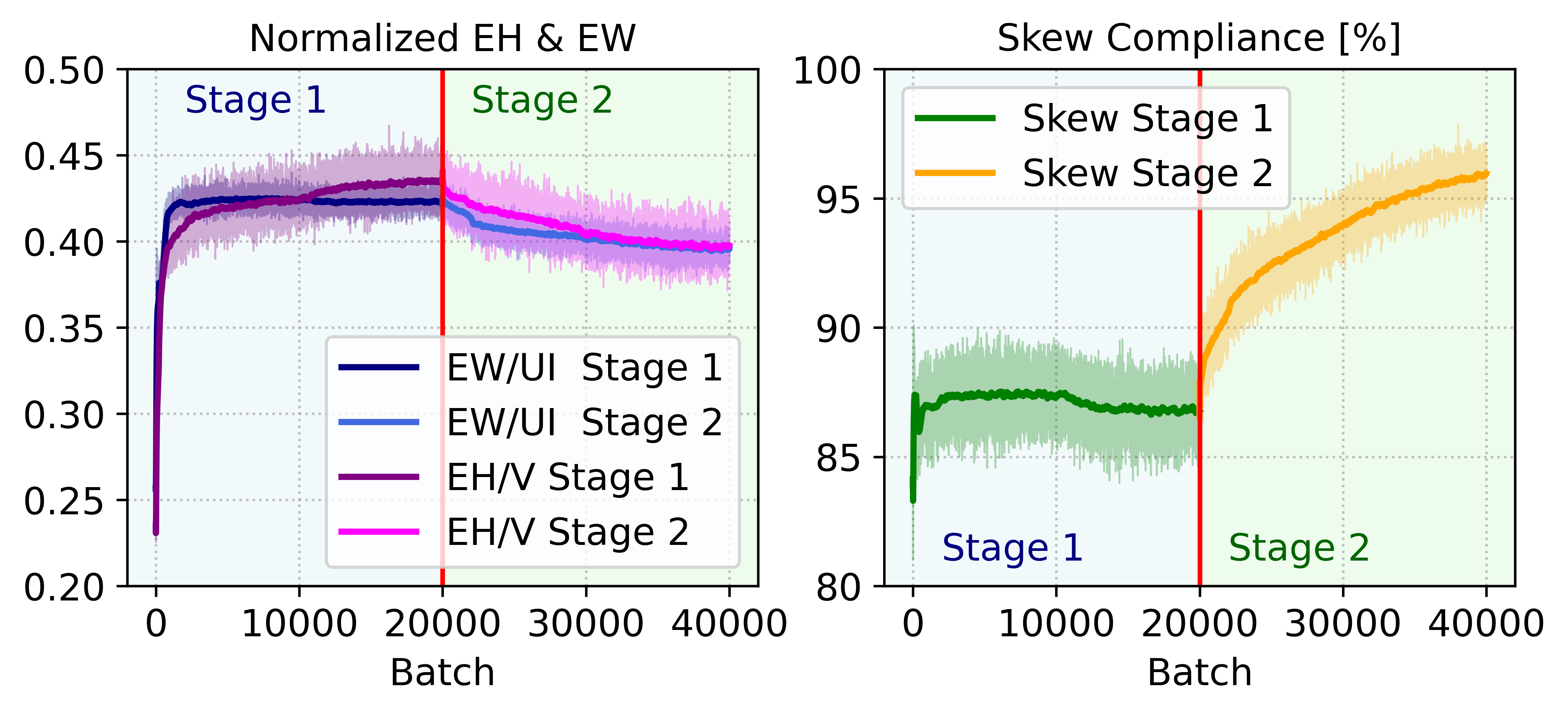}
	\caption{DDR3 DQS training process showing $EW/UI$ and $EH/V_{\mathrm{DDQ}}$ (left) as well as skew compliance rate [\%] (right).}
	\label{fig_training_ddr3_dqs}
\end{figure}

Finally, ANO is evaluated for direct constrained SI optimization instead of equalization. The scalar metrics EH, EW, and intra-pair skew were approximated using fully-connected MLP surrogate models, see Fig.~\ref{fig_ano} (b), established in \cite{Withoeft2024}. The ANO policy is learned via backpropagation through these MLPs. The near-optimal trace width $W$ and spacing $S$ as well as $R_T$ are obtained based on the design space of transmission line lengths and the dielectric constant $\epsilon_r$. Concurrently, the intra-pair skew constraints that were also given as context $C$ had to be satisfied. Compliance with these constraints based on an untrained policy network often leads to vanishing gradients or trivial solutions, e.g., closing the eye diagram to satisfy timing. Therefore, the ANO policy was trained using a two-stage curriculum training strategy, as shown in Fig.~\ref{fig_training_ddr3_dqs}.

In Stage 1, within the initial 100 epochs and 20,000 batches, the network was trained in an unconstrained warm-up phase, where only the normalized eye opening defined as $\mathcal{M}_{\mathrm{eye}} = EW/UI \cdot EH/V_{\mathrm{DDQ}}$ was maximized. As shown in Fig.~\ref{fig_training_ddr3_dqs}, the network maximizes EH and EW in this phase, but as expected, the skew compliance stayed below 90\%.

In Stage 2, within the last 100 epochs and 20,000 batches, the dynamic skew constraint was added. In every training batch and for each sample, a target skew threshold $t_{\mathrm{skew}}$ was uniformly sampled between 10 and 200~ps. This required the global ANO policy to generalize with respect to varying skew compliance limits. To ensure that there was a continuous gradient at the constraint thresholds, the total loss $\mathcal{L}$ used a differentiable Softplus penalty for skew violations:
\begin{equation}
	\mathcal{L} = -\mathcal{M}_{\mathrm{eye}} + \frac{\lambda}{\gamma}\ln\left(1+e^{\gamma(s-t_{\mathrm{skew}})}\right),
\end{equation}
where $s$ is the predicted intra-pair skew, and $\gamma=20$ controls the sharpness of the penalty. The weighting factor $\lambda$ was set to zero during Stage 1 and was increased from 0.005 to 0.05 in Stage 2. The training curves in Fig.~\ref{fig_training_ddr3_dqs} demonstrate how the network adapted to this skew penalty. As the penalty increased, skew compliance above 95\% was achieved, but this also inevitably led to a slight degradation in the absolute eye opening to satisfy the skew constraints.

\begin{table}[b!]
	\centering
	\caption{Comparison of ANO vs. GA for DDR3 DQS averaged over 25 samples per skew limit threshold}
	\label{tab_benchmark_ddr3_dqs}
	\begin{tabular}{lcccccc}
		\toprule
		\textbf{Limit} & \multicolumn{2}{c}{\textbf{Eye-Width [ps]}} & \multicolumn{2}{c}{\textbf{Eye-Height [mV]}} & \multicolumn{2}{c}{\textbf{Skew Compliance}} \\
		\cmidrule(lr){2-3} \cmidrule(lr){4-5} \cmidrule(lr){6-7}
		[ps] & \textbf{ANO} & \textbf{GA} & \textbf{ANO} & \textbf{GA} & \textbf{ANO} & \textbf{GA} \\
		\midrule
		50  & 125 & 130 & 594 & 711 & 100\% & 100\% \\
		100 & 114 & 121 & 531 & 576 & 100\% & 100\% \\
		150 & 133 & 134 & 627 & 698 & 100\% & 100\% \\
		200 & 141 & 137 & 684 & 749 & 100\% & 100\% \\
		\midrule
		\textbf{Avg.} & \textbf{128} & \textbf{131} & \textbf{609} & \textbf{684} & \textbf{100\%} & \textbf{100\%} \\
		\bottomrule
	\end{tabular}
\end{table}

To evaluate these learned constrained optimization capabilities of the trained ANO policy, it was benchmarked against the GA, which has also been utilized for feasible region optimization in \cite{Withoeft2024}. Both algorithms were tested based on four dynamic skew thresholds on 25 randomized channel contexts each, and these results are summarized in Table~\ref{tab_benchmark_ddr3_dqs}. Because the GA runs an iterative search for each specific channel instance, it achieves a marginally better EW and EH compared to ANO. The GA outperforms ANO by roughly 9.3\% on average, which corresponds to the previously discussed amortization gap. However, both methods achieve a 100\% intra-pair skew constraint compliance rate throughout all test cases. As the underlying surrogate for this task is a computationally lightweight MLP rather than the previously used more complex recurrent encoder-decoder architecture, the iterative GA is able to converge on a single channel in an absolute runtime of just 271~milliseconds on a CPU and 268~milliseconds on a GPU. The ANO framework achieves this in just 1.8~milliseconds on a standard CPU/GPU. While this GA runtime is perfectly acceptable for a single optimization run, its underlying sequential methodology is suboptimal as before when scaled to large-scale SI-DSE.

\begin{figure}[t!]
	\centering
	\includegraphics[width=8.5cm]{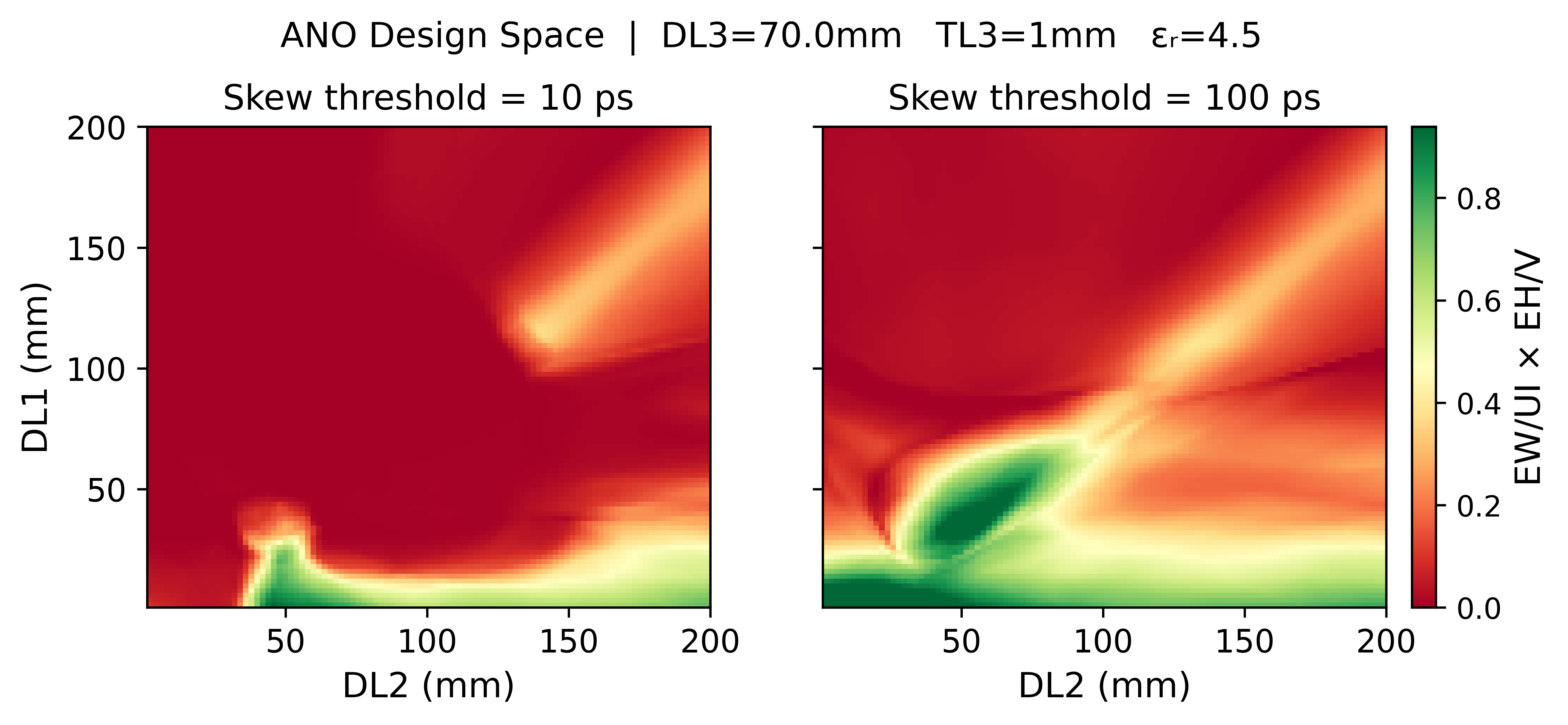}
	\caption{DDR3 DQS DSE for DL1 vs. DL2 with skew constrained to 10~ps (left) and 100~ps (right).}
	\label{fig_dse_ddr3_dqs}
\end{figure}

This scaling limitation is handled more efficiently by the ANO framework through batched inference, which transforms how SI constraints can be analyzed during pre-layout. Fig.~\ref{fig_dse_ddr3_dqs} shows a large-scale DSE where the routing lengths $DL_1$ and $DL_2$ are swept across a $100 \times 100$ grid, resulting in 10,000 design points. For every point on the grid, the ANO framework predicts the near-optimal $W$, $S$, and $R_T$ to maximize the eye under the specified skew constraint. The heatmaps show the intra-pair skew impact as with a 10~ps threshold, the feasible design space is restricted, because the optimizer faces problems when compensating for the $TL_3$ imbalance. Relaxing the constraint to 100~ps expands the valid transmission line lengths as expected. At this scale, the inefficiency of black-box search methods such as the GA becomes clear. Iterative methods scale linearly with the number of design space points, which leads to an estimated 45~minutes of sequential GA evaluations for this application. In contrast to this iterative runtime, the ANO framework solves the entire 10,000-point constrained optimization task in a single forward pass. This batched inference takes just 19~milliseconds on a CPU and 1.8~milliseconds when GPU-accelerated, which enables interactive and real-time feasibility analysis.

\section{Conclusion and Future Work}

This paper introduces an amortized neural optimization framework (ANO) for pre-layout signal integrity design space exploration. Traditional signal integrity optimization is often based on simulations and iterative black-box optimization algorithms. Even when accelerated by fast machine learning surrogates, these sequential approaches incur cumulative runtimes that are computationally expensive for large-scale multi-corner design space exploration and analysis. The proposed ANO methodology overcomes this limitation through exploitation of the continuous differentiability of neural surrogates. The framework trains a global optimization policy network offline based on surrogate gradient information. The computational cost is amortized during training, which allows the learned policy to analyze previously unseen channel contexts and provide near-optimal design configurations in a single deterministic forward pass.

The framework was evaluated using three high-speed interconnect design scenarios. For the DDR5 daisy chain topology, a multi-task ANO policy optimizes dual-receiver DFE taps based on the given channel context parameters. This eliminates the need for iterative optimization and provides a fast eye mask feasibility check to distinguish which channels can be rescued by optimal DFE settings and which remain unrecoverable. In the SerDes environment, the policy learns the synergies of 9-dimensional Tx/Rx co-equalization for a large-scale 320,000-point multi-corner analysis to obtain near-optimal EQ parameters in milliseconds. Again, verification of the predicted eye contours against a hexagonal eye mask across all manufacturing corners allows for fast compliance checking to identify which channels can still be salvaged by optimal EQ settings. Finally, the DDR3 DQS scenario demonstrates that the ANO framework also translates to constrained SI optimization using a curriculum learning strategy. The optimization target is to optimize trace geometries and termination resistors to maximize eye-height and eye-width, while also concurrently achieving compliance with intra-pair skew thresholds.

A benchmark against gradient-based, probabilistic, heuristic, as well as evolutionary and swarm optimization algorithms showed the expected trade-off. Because the global ANO policy has to generalize across the entire design space, there is an amortization gap of roughly 10\% compared to the best-performing instance-specific evolutionary and swarm algorithms. However, this small loss in absolute optimality is outweighed by the reduction in computational cost. The large-scale optimization runtimes are reduced from several hours or days down to only milliseconds. This allows the ANO framework to transform multi-corner SI analysis from a slow verification step into an interactive real-time pre-layout design space exploration tool.

Future work will focus on the integration of graph neural networks as the differentiable surrogate to allow the framework to be applied to arbitrary connectivity patterns of topologies and interconnects. In addition, future research will explore hybrid inference strategies to minimize the amortization gap. This could involve a combination of the fast ANO prediction with a fine-tuning step using gradient descent on the surrogate to recover absolute optimality for edge cases without reverting to full iterative black-box search.

\bibliography{paper_bib}

\begin{thebibliography}{10}
\providecommand{\url}[1]{#1}
\csname url@samestyle\endcsname
\providecommand{\newblock}{\relax}
\providecommand{\bibinfo}[2]{#2}
\providecommand{\BIBentrySTDinterwordspacing}{\spaceskip=0pt\relax}
\providecommand{\BIBentryALTinterwordstretchfactor}{4}
\providecommand{\BIBentryALTinterwordspacing}{\spaceskip=\fontdimen2\font plus
\BIBentryALTinterwordstretchfactor\fontdimen3\font minus
  \fontdimen4\font\relax}
\providecommand{\BIBforeignlanguage}[2]{{%
\expandafter\ifx\csname l@#1\endcsname\relax
\typeout{** WARNING: IEEEtran.bst: No hyphenation pattern has been}%
\typeout{** loaded for the language `#1'. Using the pattern for}%
\typeout{** the default language instead.}%
\else
\language=\csname l@#1\endcsname
\fi
#2}}
\providecommand{\BIBdecl}{\relax}
\BIBdecl

\bibitem{Withoeft2026}
J.~Withöft, W.~John, E.~Ecik, R.~Brüning, and J.~Götze,
  ``{Surrogate-Assisted Amortized Neural Optimization of DDR5 Decision Feedback
  Equalization},'' in \emph{2026 IEEE International Symposium on
  Electromagnetic Compatibility, Signal \& Power Integrity (EMC+SIPI)}, 2026.

\bibitem{Lu2018}
T.~Lu, J.~Sun, K.~Wu, and Z.~Yang, ``{High-Speed Channel Modeling With Machine
  Learning Methods for Signal Integrity Analysis},'' \emph{{IEEE Transactions
  on Electromagnetic Compatibility}}, vol.~60, no.~6, pp. 1957--1964, 2018.

\bibitem{Zhang2022}
H.~H. Zhang, Z.~S. Xue, X.~Y. Liu, P.~Li, L.~Jiang, and G.~M. Shi,
  ``{Optimization of High-Speed Channel for Signal Integrity With Deep Genetic
  Algorithm},'' \emph{IEEE Transactions on Electromagnetic Compatibility},
  vol.~64, no.~4, pp. 1270--1274, 2022.

\bibitem{Withoeft2024}
J.~Withöft, W.~John, E.~Ecik, R.~Brüning, and J.~Götze, ``{Machine Learning
  Methods for Elaborating the Feasible Region for Signal Integrity Analysis in
  Differential Pair PCB Structures},'' in \emph{2024 International Symposium on
  Electromagnetic Compatibility – EMC Europe}, 2024, pp. 151--156.

\bibitem{Withoeft2025}
J.~Withöft, W.~John, E.~Ecik, J.~Wastian, R.~Brüning, and J.~Götze, ``{A
  Machine Learning Modeling and Optimization Framework for Signal Integrity
  Design Support},'' in \emph{2025 Asia-Pacific International Symposium and
  Exhibition on Electromagnetic Compatibility (APEMC)}, 2025, pp. 410--413.

\bibitem{Ma2022}
H.~Ma, D.~Li, E.-P. Li, A.~C. Cangellaris, and X.~Chen, ``{A Fast Optimization
  Method for High-Speed Link Inverse Design With SVR-AS Algorithm},''
  \emph{IEEE Transactions on Signal and Power Integrity}, vol.~1, pp. 22--31,
  2022.

\bibitem{Nguyen2023}
T.~Nguyen, B.~Shi, H.~Ma, E.-P. Li, A.~Cangellaris, and J.~Schutt-Aine,
  ``{Constrained Gaussian Process for Signal Integrity applications using
  Variational Inference},'' in \emph{2023 IEEE/MTT-S International Microwave
  Symposium - IMS 2023}, 2023, pp. 155--158.

\bibitem{Goay2019}
C.~H. Goay, A.~Abd~Aziz, N.~S. Ahmad, and P.~Goh, ``{Eye Diagram Contour
  Modeling Using Multilayer Perceptron Neural Networks With Adaptive Sampling
  and Feature Selection},'' \emph{IEEE Transactions on Components, Packaging
  and Manufacturing Technology}, vol.~9, no.~12, pp. 2427--2441, 2019.

\bibitem{Telescu2023}
M.~Telescu, R.~Trinchero, N.~Soleimani, N.~Tanguy, and I.~S. Stievano,
  ``{Stochastic Time-Domain Mapping for Comprehensive Uncertainty Assessment in
  Eye Diagrams},'' \emph{IEEE Transactions on Electromagnetic Compatibility},
  vol.~65, no.~6, pp. 1930--1938, 2023.

\bibitem{Bohl2023}
L.~P. P.~B. Bohl, K.~Scharff, X.~Duan, D.~Kaller, and C.~Schuster, ``{Bayesian
  Optimization of First-Order Continuous-Time Linear Equalization in High-Speed
  Links Including Crosstalk},'' in \emph{2023 IEEE 27th Workshop on Signal and
  Power Integrity (SPI)}, 2023, pp. 1--4.

\bibitem{Dolatsara2022}
M.~Ahadi~Dolatsara, ``{A Simplified Constrained Bayesian Optimization Approach
  to Optimize the Tx Equalization in SerDes Channels},'' \emph{IEEE Letters on
  Electromagnetic Compatibility Practice and Applications}, vol.~5, no.~2, pp.
  41--47, 2023.

\bibitem{Kiguradze2020}
Z.~Kiguradze \emph{et~al.}, ``{High-Speed Channel Equalization Applying
  Parallel Bayesian Machine Learning},'' in \emph{2020 IEEE International
  Symposium on Electromagnetic Compatibility \& Signal/Power Integrity
  (EMCSI)}, 2020, pp. 251--256.

\bibitem{Dikhaminjia2021}
N.~Dikhaminjia \emph{et~al.}, ``{Optimization of Joint Equalization of
  High-Speed Signals using Bayesian Machine Learning},'' in \emph{2021 IEEE
  International Joint EMC/SI/PI and EMC Europe Symposium}, 2021, pp. 48--52.

\bibitem{Sun2024}
Z.~Sun, J.~Qiu, and H.~Ma, ``{Fast Feed-Forward Equalization Optimization
  Method for PCI Express Gen 4.0},'' in \emph{2024 International Conference on
  Microwave and Millimeter Wave Technology (ICMMT)}, vol.~1, 2024, pp. 1--3.

\bibitem{Withoeft2023a}
J.~Withöft, W.~John, E.~Ecik, R.~Brüning, and J.~Götze, ``{AI Models for
  Supporting SI Analysis on PCB Net Structures: Comparing Linear and Non-Linear
  Data Sources},'' \emph{Advances in Radio Science}, vol.~21, pp. 77--87, 2023.

\bibitem{Withoeft2023b}
J.~Withöft, W.~John, E.~Ecik, and J.~Götze, ``{AI-Based SI-Compliant PCB
  Design Support for DDR Technology Enhanced by Transfer Learning},'' in
  \emph{2023 International Symposium on Electromagnetic Compatibility – EMC
  Europe}, 2023, pp. 1--6.

\bibitem{Trinchero2019}
R.~Trinchero, M.~A. Dolatsara, K.~Roy, M.~Swaminathan, and F.~G. Canavero,
  ``{Design of High-Speed Links via a Machine Learning Surrogate Model for the
  Inverse Problem},'' in \emph{2019 Electrical Design of Advanced Packaging and
  Systems (EDAPS)}, 2019, pp. 1--3.

\bibitem{Roy2019}
K.~Roy, M.~A. Dolatsara, H.~M. Torun, R.~Trinchero, and M.~Swaminathan,
  ``{Inverse Design of Transmission Lines with Deep Learning},'' in \emph{{2019
  IEEE 28th Conference on Electrical Performance of Electronic Packaging and
  Systems (EPEPS)}}, 2019, pp. 1--3.

\bibitem{Dolatsara2020}
M.~A. Dolatsara, H.~Yu, J.~A. Hejase, W.~Dale~Becker, and M.~Swaminathan,
  ``{Invertible Neural Networks for Inverse Design of CTLE in High-speed
  Channels},'' in \emph{2020 IEEE Electrical Design of Advanced Packaging and
  Systems (EDAPS)}, 2020, pp. 1--3.

\bibitem{Ambasana2021}
N.~Ambasana \emph{et~al.}, ``{Invertible Neural Networks for High-Speed Channel
  Design \& Parameter Distribution Estimation},'' in \emph{2021 IEEE 30th
  Conference on Electrical Performance of Electronic Packaging and Systems
  (EPEPS)}, 2021, pp. 1--3.

\bibitem{Bhatti2021}
O.~W. Bhatti \emph{et~al.}, ``{Comparison of Invertible Architectures for High
  Speed Channel Design},'' in \emph{2021 IEEE Electrical Design of Advanced
  Packaging and Systems (EDAPS)}, 2021, pp. 1--3.

\bibitem{Lho2022}
D.~Lho \emph{et~al.}, ``{Deterministic Policy Gradient-based Reinforcement
  Learning for DDR5 Memory Signaling Architecture Optimization considering
  Signal Integrity},'' in \emph{2022 IEEE 31st Conference on Electrical
  Performance of Electronic Packaging and Systems (EPEPS)}, 2022, pp. 1--3.

\bibitem{Choi2023}
S.~Choi \emph{et~al.}, ``{Deep Reinforcement Learning-Based Optimal and Fast
  Hybrid Equalizer Design Method for High-Bandwidth Memory (HBM) Module},''
  \emph{IEEE Transactions on Components, Packaging and Manufacturing
  Technology}, vol.~13, no.~11, pp. 1804--1816, 2023.

\bibitem{Withoeft2023c}
J.~Withöft, W.~John, E.~Ecik, R.~Brüning, and J.~Götze, ``{Optimization of a
  Daisy Chain PCB Memory System through Reinforcement Learning under
  Consideration of Signal Integrity Constraints},'' in \emph{2023 Kleinheubach
  Conference}, 2023, pp. 1--4.

\bibitem{Kim2023a}
J.~Kim \emph{et~al.}, ``{Imitation Learning-based Equalizer Design Optimization
  Method on PCIe 6.0},'' in \emph{2023 IEEE Electrical Design of Advanced
  Packaging and Systems (EDAPS)}, 2023, pp. 1--3.

\bibitem{Choi2024}
S.~Choi \emph{et~al.}, ``{Imitation Learning-based Fast Optimization of SSD
  Interface for PCIe 6.0 considering Signal Integrity},'' in \emph{2024 IEEE
  International Symposium on Electromagnetic Compatibility, Signal \& Power
  Integrity (EMC+SIPI)}, 2024, pp. 1--6.

\bibitem{Kim2023b}
J.~Kim \emph{et~al.}, ``{Bayesian Exploration Imitation Learning-Based
  Contextual via Design Optimization Method of PAM-4-Based High-Speed Serial
  Link},'' \emph{IEEE Transactions on Electromagnetic Compatibility}, vol.~65,
  no.~6, pp. 1751--1762, 2023.

\bibitem{Akinwande2023}
O.~Akinwande, S.~Erdogan, R.~Kumar, and M.~Swaminathan, ``{Surrogate Modeling
  With Complex-Valued Neural Nets for Signal Integrity Applications},''
  \emph{IEEE Transactions on Microwave Theory and Techniques}, vol.~72, no.~1,
  pp. 478--489, 2024.

\bibitem{Sun2021}
X.~Sun, T.~Xue, S.~Rusinkiewicz, and R.~P. Adams, ``{Amortized Synthesis of
  Constrained Configurations Using a Differentiable Surrogate},'' in
  \emph{Advances in Neural Information Processing Systems (NeurIPS)}, vol.~34,
  2021, pp. 18\,891--18\,906.

\bibitem{Amos2023}
B.~Amos, ``{Tutorial on Amortized Optimization},'' \emph{Foundations and
  Trends® in Machine Learning}, vol.~16, no.~5, pp. 592--732, 2023.

\bibitem{jedec}
\BIBentryALTinterwordspacing
{JEDEC Solid State Technology Association}, ``{DDR5 SDRAM Standard},'' JEDEC,
  Arlington, VA, USA, Standard JESD79-5D, 2025. [Online]. Available:
  \url{https://www.jedec.org/standards-documents/docs/jesd79-5d}
\BIBentrySTDinterwordspacing

\bibitem{LTspice}
\BIBentryALTinterwordspacing
{Analog Devices}, ``{LTspice XVII},'' 2016. [Online]. Available:
  \url{https://www.analog.com/en/design-center/design-tools-and-calculators/ltspice-simulator.html}
\BIBentrySTDinterwordspacing

\bibitem{Matlab}
\BIBentryALTinterwordspacing
{Mathworks}, ``{Matlab Version R2025b},'' 2025. [Online]. Available:
  \url{https://de.mathworks.com/products/matlab.html}
\BIBentrySTDinterwordspacing

\bibitem{eCADSTAR}
\BIBentryALTinterwordspacing
{Zuken}, ``{eCADSTAR Version 2021.1},'' 2021. [Online]. Available:
  \url{https://www.ecadstar.com/en}
\BIBentrySTDinterwordspacing

\bibitem{Goldberg1989}
D.~E. Goldberg, \emph{{Genetic Algorithms in Search, Optimization, and Machine
  Learning}}.\hskip 1em plus 0.5em minus 0.4em\relax Reading, MA, USA:
  Addison-Wesley, 1989.

\bibitem{Storn1997}
R.~Storn and K.~V. Price, ``{Differential Evolution – A Simple and Efficient
  Heuristic for global Optimization over Continuous Spaces},'' \emph{Journal of
  Global Optimization}, vol.~11, pp. 341--359, 1997.

\bibitem{Kennedy1995}
J.~Kennedy and R.~Eberhart, ``{Particle swarm optimization},'' in
  \emph{Proceedings of ICNN'95 - International Conference on Neural Networks},
  vol.~4, 1995, pp. 1942--1948.

\bibitem{Hansen2001}
N.~Hansen and A.~Ostermeier, ``{Completely Derandomized Self-Adaptation in
  Evolution Strategies},'' \emph{Evolutionary Computation}, vol.~9, no.~2, pp.
  159--195, 2001.

\bibitem{Mockus1974}
J.~Mockus, ``{On Bayesian Methods for Seeking the Extremum},'' in
  \emph{Proceedings of the IFIP Technical Conference}, 1974, pp. 400--404.

\bibitem{Nelder1965}
J.~A. Nelder and R.~Mead, ``{A Simplex Method for Function Minimization},''
  \emph{The Computer Journal}, vol.~7, no.~4, pp. 308--313, 1965.

\end{thebibliography}

\bibliographystyle{IEEEtran}

\clearpage

\onecolumn

\appendices
\section{Additional SerDes Design Space Exploration Results}
\label{cha_appendix}
\vspace{0.5cm}

\begin{figure}[H]
	\centering
	\includegraphics[width=\textwidth]{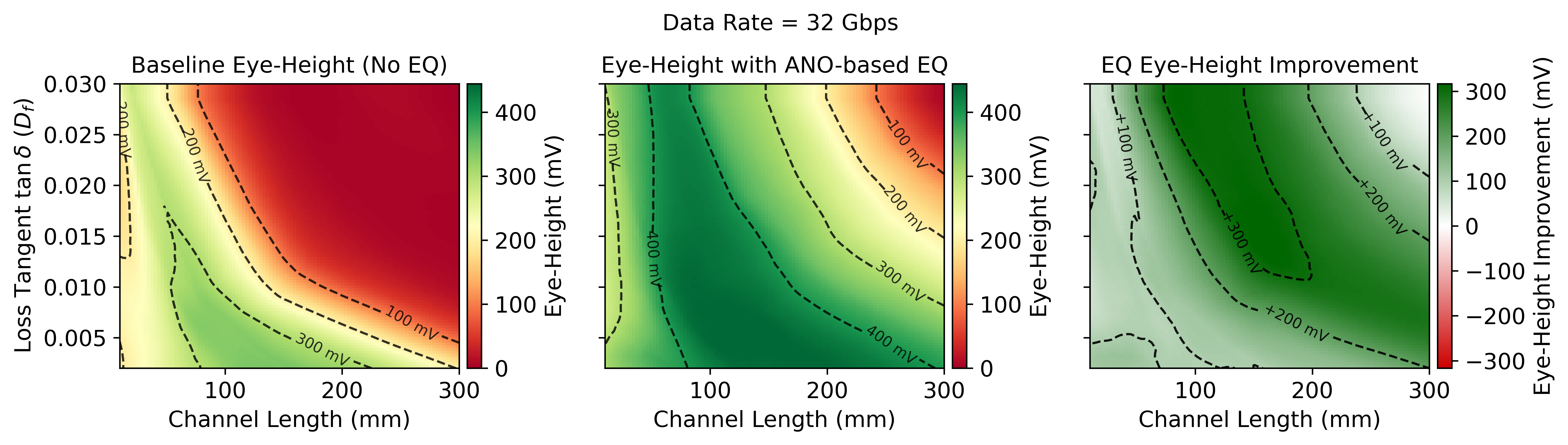}
	\caption{Large-scale design space exploration example using the ANO-based EQ and surrogate predictions of worst-case EH performance over a $100 \times 100$ grid of channel lengths and loss tangents at 32 manufacturing corner cases and a nominal data rate of 32~Gbps for the unequalized baseline (left), ANO-based EQ (center), and ANO-based EQ EH improvement (right).}
	\label{fig_dse_losstan_vs_length_serdes}
\end{figure}

\begin{figure}[H]
	\centering
	\includegraphics[width=\textwidth]{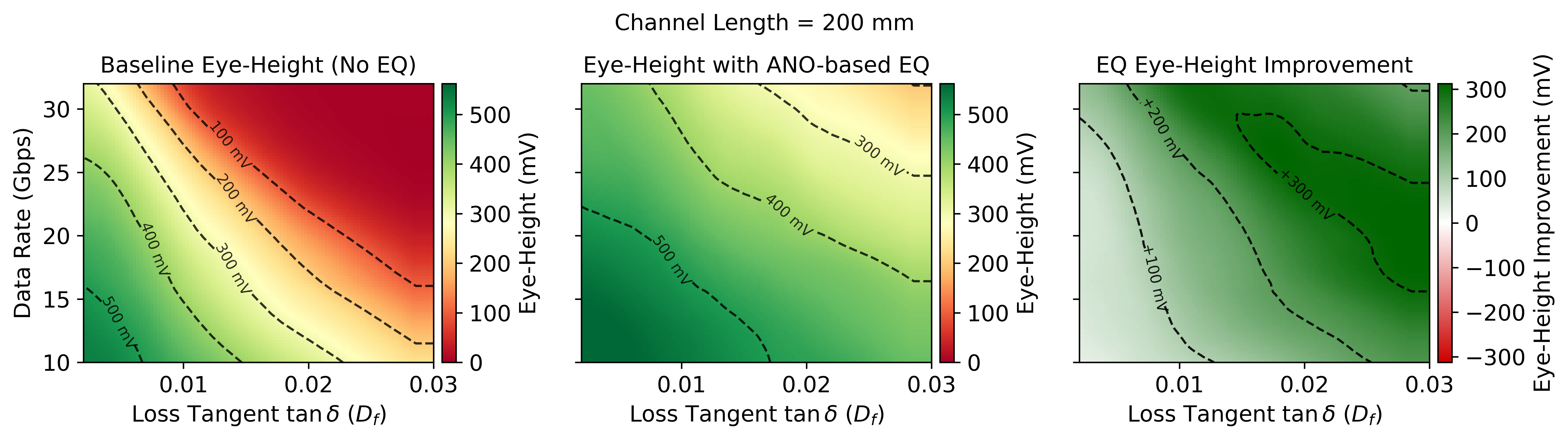}
	\caption{Large-scale design space exploration example using the ANO-based EQ and surrogate predictions of worst-case EH performance over a $100 \times 100$ grid of loss tangents and data rates at 32 manufacturing corner cases and a nominal channel length ($L=200$~mm) for the unequalized baseline (left), ANO-based EQ (center), and ANO-based EQ EH improvement (right).}
	\label{fig_dse_datarate_vs_losstan_serdes}
\end{figure}

\begin{figure}[H]
	\centering
	\includegraphics[width=\textwidth]{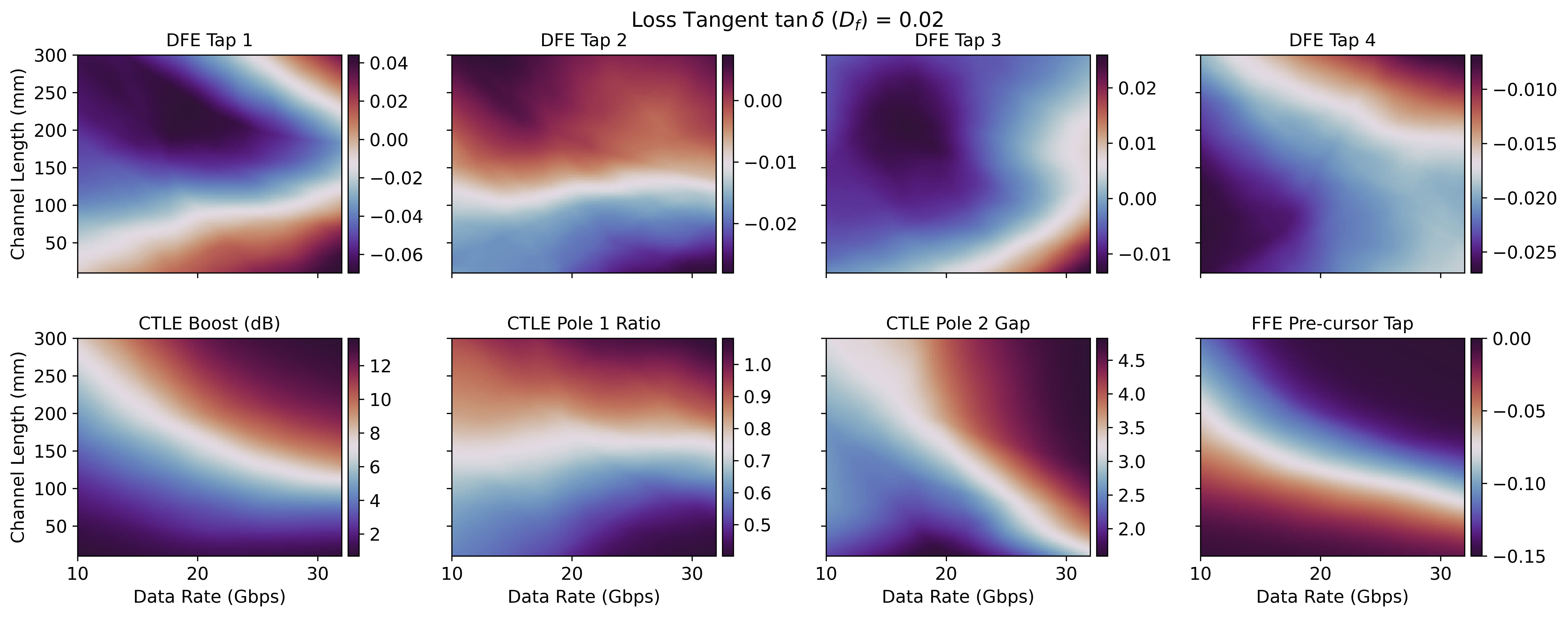}
	\caption{Equalization parameters for the nominal $W$, $S$, $H$, $D_k$, and $D_f$ geometry parameters across the $100 \times 100$ grid of data rates and channel lengths.}
	\label{fig_dse_eq_length_vs_datarate_serdes}
\end{figure}

\clearpage

\begin{figure}[H]
	\centering
	\includegraphics[width=\textwidth]{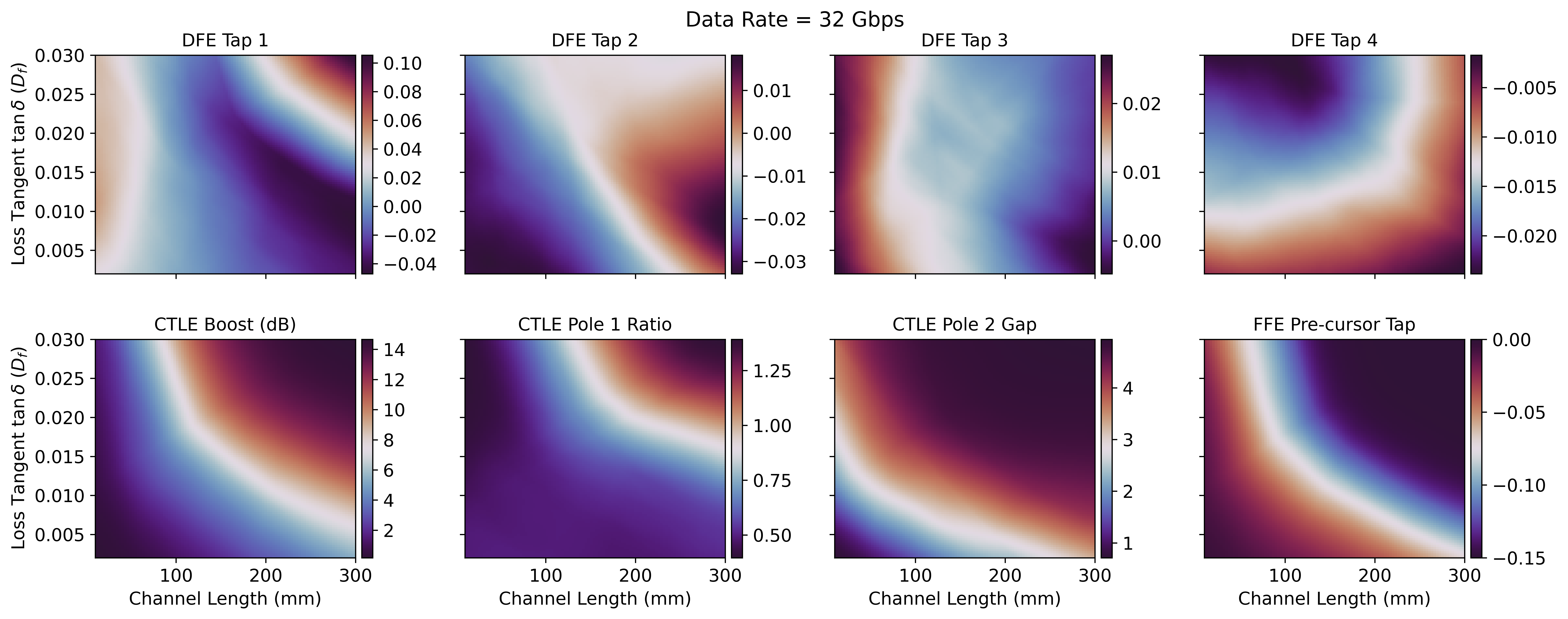}
	\caption{Equalization parameters for the nominal $W$, $S$, $H$, $D_k$, and $D_f$ geometry parameters across the $100 \times 100$ grid of channel lengths and loss tangents.}
	\label{fig_dse_eq_losstan_vs_length_serdes}
\end{figure}

\begin{figure}[H]
	\centering
	\includegraphics[width=\textwidth]{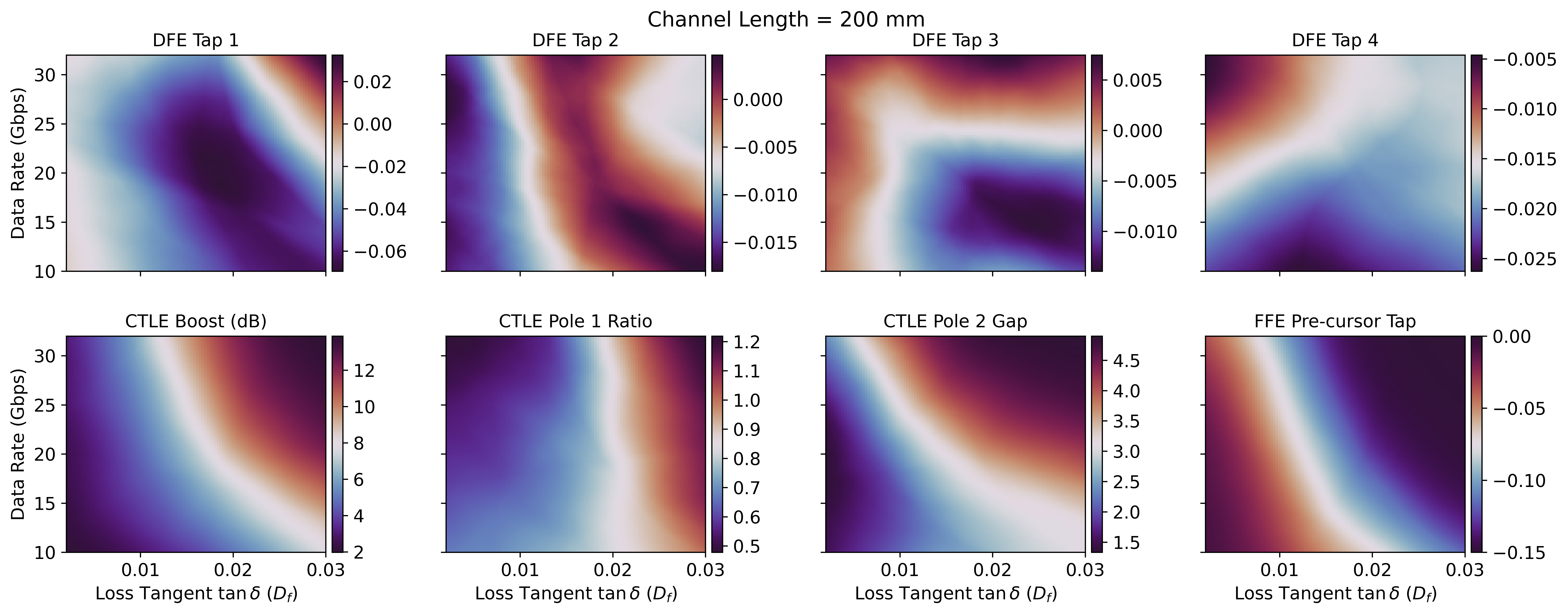}
	\caption{Equalization parameters for the nominal $W$, $S$, $H$, $D_k$, and $D_f$ geometry parameters across the $100 \times 100$ grid of loss tangents and data rates.}
	\label{fig_dse_eq_datarate_vs_losstan_serdes}
\end{figure}

\end{document}